\documentclass[twocolumn,canadian]{svjour3N}
\usepackage[latin9]{inputenc}

\usepackage{color}
\usepackage{babel}
\usepackage{array}
\usepackage{rotating}
\usepackage{amsmath}
\usepackage{amssymb}
\usepackage{graphicx}
\usepackage{setspace}
\usepackage[authoryear]{natbib}
\usepackage{nicefrac}
\usepackage{xargs}[2008/03/08]
\usepackage[unicode=true,pdfusetitle,
 bookmarks=true,bookmarksnumbered=false,bookmarksopen=false,
 breaklinks=true,pdfborder={0 0 0},pdfborderstyle={},backref=false,colorlinks=true]
 {hyperref}
\hypersetup{
 linkcolor=pigment, citecolor=pigment, urlcolor=pigment}

\makeatletter

\providecommand{\tabularnewline}{\\}

\definecolor{pigment}{rgb}{0.2, 0.2, 0.6}
\usepackage{arydshln}
\usepackage{breakcites}
\usepackage{amsfonts}

\@ifundefined{showcaptionsetup}{}{%
 \PassOptionsToPackage{caption=false}{subfig}}
\usepackage{subfig}
\makeatother

\begin{document}
\sloppy

\global\long\def\argparentheses#1{\mathopen{\left(#1\right)}\mathclose{}}%
\global\long\def\ap#1{\argparentheses{#1}\mathclose{}}%

\global\long\def\var#1{\mathbb{V}\argparentheses{#1}}%
\global\long\def\pr#1{\mathbb{P}\argparentheses{#1}}%
\global\long\def\ev#1{\mathbb{E}\argparentheses{#1}}%

\global\long\def\smft#1#2{\stackrel[#1]{#2}{\sum}}%
\global\long\def\smo#1{\underset{#1}{\sum}}%
\global\long\def\prft#1#2{\stackrel[#1]{#2}{\prod}}%
\global\long\def\pro#1{\underset{#1}{\prod}}%
\global\long\def\uno#1{\underset{#1}{\bigcup}}%

\global\long\def\order#1{\mathcal{O}\argparentheses{#1}}%
\global\long\def\R{\mathbb{R}}%
\global\long\def\Q{\mathbb{Q}}%
\global\long\def\N{\mathbb{N}}%
\global\long\def\F{\mathcal{F}}%

\global\long\def\mathtext#1{\mathrm{#1}}%

\global\long\def\maxo#1{\underset{#1}{\max\,}}%
\global\long\def\argmaxo#1{\underset{#1}{\mathtext{argmax}\,}}%
\global\long\def\argsupo#1{\underset{#1}{\mathtext{argsup}\,}}%
\global\long\def\supo#1{\underset{#1}{\sup\,}}%
\global\long\def\info#1{\underset{#1}{\inf\,}}%
\global\long\def\mino#1{\underset{#1}{\min\,}}%
\global\long\def\argmino#1{\underset{#1}{\mathtext{argmin}\,}}%
\global\long\def\limo#1#2{\underset{#1\rightarrow#2}{\lim}}%
\global\long\def\supo#1{\underset{#1}{\sup}}%
\global\long\def\info#1{\underset{#1}{\inf}}%

\global\long\def\b#1{\boldsymbol{#1}}%
\global\long\def\ol#1{\overline{#1}}%
\global\long\def\ul#1{\underline{#1}}%

\newcommandx\der[3][usedefault, addprefix=\global, 1=, 2=, 3=]{\frac{d^{#2}#3}{d#1^{#2}}}%
\newcommandx\deri[3][usedefault, addprefix=\global, 1=, 2=, 3=]{\nicefrac{d^{#2}#3}{d#1^{#2}}}%
\newcommandx\pderi[3][usedefault, addprefix=\global, 1=, 2=]{\nicefrac{\partial^{#2}#3}{\partial#1^{#2}}}%
\newcommandx\pder[3][usedefault, addprefix=\global, 1=, 2=]{\frac{\partial^{#2}#3}{\partial#1^{#2}}}%
\global\long\def\intft#1#2#3#4{\int\limits _{#1}^{#2}#3d#4}%
\global\long\def\into#1#2#3{\underset{#1}{\int}#2d#3}%

\global\long\def\th{\theta}%
\global\long\def\la#1{~#1~}%
\global\long\def\laq{\la =}%
\global\long\def\normal#1#2{\mathcal{N}\argparentheses{#1,\,#2}}%
\global\long\def\uniform#1#2{\mathcal{U}\argparentheses{#1,\,#2}}%
\global\long\def\I#1#2{\mbox{I}_{#1}\argparentheses{#2}}%
\global\long\def\chisq#1{\chi_{#1}^{2}}%
\global\long\def\dar{\,\Longrightarrow\,}%
\global\long\def\dal{\,\Longleftarrow\,}%
\global\long\def\dad{\,\Longleftrightarrow\,}%
\global\long\def\norm#1{\left\Vert #1\right\Vert }%
\global\long\def\code#1{\mathtt{#1}}%
\global\long\def\descr#1#2{\underset{#2}{\underbrace{#1}}}%
\global\long\def\NB{\mathcal{NB}}%
\global\long\def\BNB{\mathcal{BNB}}%
\global\long\def\e#1{\text{e}_{#1}}%
\global\long\def\P{\mathbb{P}}%
\global\long\def\pb#1{\Bigg(#1\Bigg)}%
\global\long\def\cpb#1{\Big[#1\Big]}%

\global\long\def\vv#1{\boldsymbol{#1}}%

\global\long\def\mt#1{\mathtext{#1}}%

\global\long\def\mat#1{\mt{#1}}%

\global\long\def\mm#1{\underline{\boldsymbol{#1}}}%

\global\long\def\G{g}%

\global\long\def\Gv{\vv{\G}}%

\global\long\def\H{\mat H}%

\global\long\def\Hv{\vv{\mat{\H}}}%

\global\long\def\Hm{\mm{\H}}%

\global\long\def\pl{\ell_{\mt{PL}}}%

\global\long\def\dx{\vv{\delta}}%
\global\long\def\dxt{\widetilde{\dx}}%
\global\long\def\dxn{\delta}%

\global\long\def\wt#1{\widetilde{#1}}%

\title{A Robust and Efficient Algorithm to Find Profile Likelihood Confidence
Intervals}

\author{Samuel M. Fischer         \and
		Mark A. Lewis
}


\institute{S. M. Fischer \at
	Department of Mathematical and Statistical Sciences, 632 Central Academic Building, University of Alberta, Edmonton, AB, T6G 2G1.
	\email{samuel.fischer@ualberta.ca}           
	\and
	M. A. Lewis \at
	Department of Mathematical and Statistical Sciences, 632 Central Academic Building, University of Alberta, Edmonton, AB, T6G 2G1. \\
	Department of Biological Sciences, CW 405, Biological Sciences Building, University of Alberta, Edmonton, AB, T6G 2E9.
}

\date{}

\maketitle

\begin{abstract}
Profile likelihood confidence intervals are a robust alternative to
Wald\textquoteright s method if the asymptotic properties of the maximum
likelihood estimator are not met. However, the constrained optimization
problem defining profile likelihood confidence intervals can be difficult
to solve in these situations, because the likelihood function may
exhibit unfavorable properties. As a result, existing methods may
be inefficient and yield misleading results. In this paper, we address
this problem by computing profile likelihood confidence intervals
via a trust-region approach, where steps computed based on local approximations
are constrained to regions where these approximations are sufficiently
precise. As our algorithm also accounts for numerical issues arising
if the likelihood function is strongly non-linear or parameters are
not estimable, the method is applicable in many scenarios where earlier
approaches are shown to be unreliable. To demonstrate its potential
in applications, we apply our algorithm to benchmark problems and
compare it with 6 existing approaches to compute profile likelihood
confidence intervals. Our algorithm consistently achieved higher success
rates than any competitor while also being among the quickest methods.
As our algorithm can be applied to compute both confidence intervals
of parameters and model predictions, it is useful in a wide range
of scenarios.
\end{abstract}

\vspace*{\fill}

\section*{Declarations}

\paragraph{Funding.}
SMF is thankful for the funding received from the Canadian
Aquatic Invasive Species Network and the Natural Sciences and Engineering
Research Council of Canada (NSERC); MAL gratefully acknowledges an
NSERC Discovery Grant and Canada Research Chair.

\paragraph{Competing interests.}
The authors declare no competing interests.

\paragraph{Availability of data and material.} 
Not applicable.

\paragraph{Code availability.}
 A Python implementation of the described algorithm
and the test procedures can be retrieved from the python package index as
package \mbox{\emph{ci-rvm}} (see \href{https://pypi.org/project/ci-rvm/}{pypi.org/project/ci-rvm}).

\paragraph{Authors' contributions.} 
Samuel M. Fischer and Mark A. Lewis jointly conceived the project; Samuel M. Fischer conceived the algorithm, conducted the mathematical analysis, implemented the algorithm, and wrote the manuscript. Mark A. Lewis revised the manuscript.

\begin{acknowledgements}
	The authors would like to give thanks to the members of the Lewis
	Research Group at the University of Alberta for helpful feedback and
	discussions. 
\end{acknowledgements}

\section{Introduction}

\subsection{Profile likelihood confidence intervals}

Confidence intervals are an important tool for statistical inference,
used not only to assess the range of predictions that are supported
by a model and data but also to detect potential estimability issues
\citep{raue_structural_2009}. These estimability issues occur if
the available data do not suffice to infer a statistical quantity
on the desired confidence level, and the corresponding confidence
intervals are infinite \citep{raue_structural_2009}. Due to the broad
range of applications, confidence intervals are an integral part of
statistical model analysis and widely used across disciplines.

Often, confidence intervals are constructed via Wald's method, which
exploits the asymptotic normality of the maximum likelihood estimator
(MLE). Though Wald's method is accurate in ``benign'' use cases,
the approach can be imprecise or fail if not enough data are available
to reach the asymptotic properties of the MLE. This will be the case,
in particular, if the MLE is not unique, i.e. parameters are not identifiable,
or if the likelihood is very sensitive to parameter changes beyond
some threshold, e.g. in dynamical systems undergoing bifurcations.
Therefore, other methods, such as profile likelihood techniques \citep{cox_analysis_1989},
are favorable in many use cases.

Both Wald-type and profile likelihood confidence intervals are constructed
by inverting the likelihood likelihood ratio test. That is, the confidence
interval for a parameter $\th_{0}$ encompasses all values $\bar{\th}_{0}$
that might suit as acceptable null hypotheses if the parameter were
to be fixed; i.e. $H_{0}:\th_{0}=\bar{\th}_{0}$ could not be rejected
versus the alternative $H_{1}:\th_{0}\neq\bar{\th}_{0}$. As the likelihood
ratio statistic is, under regularity conditions, approximately $\chisq{}$
distributed under the null hypothesis, the confidence interval is
given by
\begin{eqnarray}
\mathit{I} & = & \left[\bar{\th}_{0}\,\Big|\,2\left(\maxo{\vv{\th}\in\Theta}\ell\ap{\vv{\th}}-\maxo{\vv{\th}\in\Theta\,:\,\th_{0}=\bar{\th}_{0}}\ell\ap{\vv{\th}}\right)\leq\chi_{1,1-\alpha}^{2}\right],\label{eq:base-CI}
\end{eqnarray}
whereby $\Theta$ is the parameter space, $\ell$ denotes the log-likelihood
function, $\alpha$ is the desired confidence level, and $\chi_{k,1-\alpha}^{2}$
is the $\left(1-\alpha\right)$th quantile of the $\chisq{}$ distribution
with $k$ degrees of freedom.

The function that maps $\bar{\th}_{0}$ to the constrained maximum
\begin{eqnarray}
\pl\ap{\bar{\th}_{0}} & := & \maxo{\vv{\th}\in\Theta\,:\,\th_{0}=\bar{\th}_{0}}\ell\ap{\vv{\th}}\label{eq:profile-liekllhood}
\end{eqnarray}
is called the profile log-likelihood. While Wald's method approximates
$\ell$ and $\pl$ as quadratic functions, profile likelihood confidence
intervals are constructed by exact computation of the profile log-likelihood
$\pl$. This makes this method more accurate but also computationally
challenging.

\subsection{Existing approaches}

Conceptually, the task of identifying the end points $\th_{0}^{\min}$
and $\th_{0}^{\max}$ of the confidence interval $I$ is equivalent
to finding the maximal (or minimal) value for $\th_{0}$ with 
\begin{equation}
\pl\ap{\th_{0}}\laq\ell^{*}\la{:=}\ell\ap{\hat{\vv{\th}}}-\frac{1}{2}\chi_{1,1-\alpha}^{2},
\end{equation}
Here, $\hat{\vv{\th}}$ denotes the MLE; the value $\ell^{*}$ follows
from rearranging the terms in the inequality characterizing $I$ (see
equation (\ref{eq:base-CI})).

There are two major perspectives to address this problem. It could
either be understood as a one-dimensional root finding problem on
$\pl$ or as the constrained maximization (or minimization) problem
\begin{eqnarray}
\th_{0}^{\max} & = & \maxo{\vv{\th}\in\Theta\,:\,\ell\ap{\vv{\th}}\geq\ell^{*}}\th_{0}\label{eq:constrained-max}
\end{eqnarray}
($\th_{0}^{\min}$ analog). Approaches developed from either perspective
face the challenge of balancing robustness against efficiency.

The root finding perspective \citep{cook_confidence_1990,diciccio_implementation_1991,stryhn_confidence_2003,moerbeek_comparison_2004,ren_algorithm_2019}
is robust if small steps are taken and solutions of the maximization
problem (\ref{eq:profile-liekllhood}) are good initial guesses for
the maximizations in later steps. Nonetheless, the step size should
be variable if parameters might be not estimable and the confidence
intervals large. At the same time, care must be taken with large steps,
as solving (\ref{eq:profile-liekllhood}) can be difficult if the
initial guesses are poor, and algorithms may fail to converge. Therefore,
conservative step choices are often advisable even though they may
decrease the overall efficiency of the approaches.

The constrained maximization perspective \citep{neale_use_1997,wu_adjusted_2012}
has the advantage that efficient solvers for such problems are readily
implemented in many optimization packages. If the likelihood function
is ``well behaved'', these methods converge very quickly. However,
in practical problems, the likelihood function may have local extrema,
e.g. due to lack of data, or steep ``cliffs'' that may hinder these
algorithms from converging to a feasible solution. Furthermore, general
algorithms are typically not optimized for problems like (\ref{eq:constrained-max}),
in which the target function is simple and the major challenge is
in ensuring that the constraint is met. Therefore, an approach would
be desirable that is specifically tailored to solve the constrained
maximization (\ref{eq:constrained-max}) in a robust and efficient
manner.

A first step in this direction is the algorithm by \citet{venzon_method_1988},
which solves (\ref{eq:constrained-max}) by repeated quadratic approximations
of the likelihood surface. As the method is of Newton-Raphson type,
it is very efficient as long as the local approximations are accurate.
Therefore, the algorithm is fast if the asymptotic normality of the
MLE is achieved approximately. Otherwise, the algorithm relies heavily
on good initial guesses. Though methods to determine accurate initial
guesses exist \citep{gimenez_efficient_2005}, the algorithm by \citet{venzon_method_1988}
(below abbreviated as VM) can get stuck in local extrema or fail to
converge if the likelihood surface has unfavorable properties \citep[see e.g.][]{ren_algorithm_2019}.
Moreover, the algorithm will break down if parameters are not identifiable.
Thus, VM cannot be applied in important use cases of profile likelihood
confidence intervals.

\subsection{Our contributions}

In this paper, we address the issues of VM by introducing an algorithm
extending the ideas of \citet{venzon_method_1988}. Our algorithm,
which we will call \emph{Robust Venzon-Moolgavkar Algorithm} (RVM)
below, combines the original procedure with a trust region approach
\citep{conn_trust-region_2000}. That is, the algorithm never steps
outside of the region in which the likelihood approximation is sufficiently
precise. Furthermore, RVM accounts for unidentifiable parameters,
local minima and maxima, and sharp changes in the likelihood surface.
Though RVM may not outcompete traditional approaches in problems with
well-behaved likelihood functions or in the absence of estimability
issues, we argue that RVM is a valuable alternative in the (common)
cases that the likelihood function is hard to optimize and the model
involves parameters that are not estimable.

Another well-known limitation of the approach by \citet{venzon_method_1988}
is that it is not directly applicable to construct confidence intervals
for functions of parameters. Often the main research interest is not
in identifying specific model parameters but in obtaining model predictions,
which can be expressed as a function of the parameters. In addition
to presenting a robust algorithm to find confidence intervals for
model parameters, we show how RVM (and the original VM) can also be
applied to determine confidence intervals for functions of parameters.

This paper is structured as follows: in the first section, we start
by outlining the main ideas behind RVM before we provide details of
the applied procedures. Furthermore, we briefly describe how the algorithm
can be used to determine confidence intervals of functions of parameters.
In the second section, we apply RVM and alternative algorithms to
benchmark problems with simulated data. Thereby, we review the implemented
alternative algorithms before we present the results. We conclude
this paper with a discussion of the benchmark results and the benefits
and limitations of RVM in comparison to earlier methods.

All code used in this study, including a Python implementation of
RVM, can be retrieved from the python package index as
package \mbox{\emph{ci-rvm}} (see \href{https://pypi.org/project/ci-rvm/}{pypi.org/project/ci-rvm}).

\section{Algorithm}

\subsection{Basic ideas \label{subsec:Basic-idea-of-algorithm}}

Suppose we consider a model with an $n$-dimensional parameter vector
$\vv{\th}:=\left(\th_{0},\dots,\th_{n-1}\right)$ and a twice continuously
differentiable log-likelihood function $\ell$. Assume without loss
of generality that we seek to construct a level-$\alpha$ confidence
interval for the parameter $\th_{0}$, and let $\wt{\vv{\th}}:=\left(\th_{1},\dots,\th_{n-1}\right)^{\top}$
be the vector of all remaining parameters, called nuisance parameters.
For convenience, we may write $\ell=\ell\ap{\vv{\th}}$ as a function
of the complete parameter vector or $\ell=\ell\left(\th_{0},\wt{\vv{\th}}\right)$
as a function of the parameter of interest and the nuisance parameters.

The algorithm RVM introduced in this paper searches the right end
point $\th_{0}^{\max}$ (equation (\ref{eq:constrained-max})) of
the confidence interval $\mathinner{I}$. The left end point can be
identified with the same approach if a modified model is considered
in which $\ell$ is flipped in $\th_{0}$. As RVM builds on the method
by \citet{venzon_method_1988}, we start by recapitulating their algorithm
VM below.

Let $\vv{\th}^{*}\in\Theta$ be the parameter vector at which the
parameter of interest is maximal, $\th_{0}^{*}=\th_{0}^{\max}$, and
$\ell\ap{\vv{\th}^{*}}\geq\ell^{*}$. \citet{venzon_method_1988}
note that $\vv{\th}^{*}$ satisfies the following necessary conditions:
\begin{enumerate}
\item $\ell\ap{\vv{\th}^{*}}=\ell^{*}$ and \label{enu:cond-val}
\item $\ell$ is in a local maximum with respect to the nuisance parameters,
which implies $\pderi[\wt{\vv{\th}}]{\ell}\ap{\vv{\th}^{*}}=0$.\label{enu:cond-grad}
\end{enumerate}

The algorithm VM searches for $\vv{\th}^{*}$ by minimizing both the
log-likelihood distance to the threshold $\left|\ell(\vv{\th})-\ell^{*}\right|$
and the gradient of the nuisance parameters $\pderi[\wt{\vv{\th}}]{\ell}$.
To this end, the algorithm repeatedly approximates the log-likelihood
surface $\ell$ with second order Taylor expansions $\hat{\ell}$.
If $\vv{\th}^{(i)}$ is the parameter vector in the $i^{\text{th}}$
iteration of the algorithm, expanding $\ell$ around $\vv{\th}^{(i)}$
yields
\begin{eqnarray}
\hat{\ell}(\vv{\th}) & := & \ell\ap{\vv{\th}^{(i)}}+\Gv^{\top}\left(\vv{\th}-\vv{\th}^{(i)}\right) \nonumber \\
& & +\frac{1}{2}\left(\vv{\th}-\vv{\th}^{(i)}\right)^{\top}\mm{\H}\left(\vv{\th}-\vv{\th}^{(i)}\right) \nonumber \\
 & = & \bar{\ell}+\widetilde{\Gv}^{\top}\widetilde{\dx}+\G_{0}\dxn_{0}+\frac{1}{2}\widetilde{\dx}^{\top}\widetilde{\mm{\H}}\widetilde{\dx}+\dxn_{0}\widetilde{\vv{\H}}_{0}^{\top}\widetilde{\dx} +\frac{1}{2}\dxn_{0}\H_{00}\dxn_{0}\la\nonumber \\
 & {=:} &   \hat{\ell}^{\dxn}\left(\dxn_{0},\dxt\right).\label{eq:approx-l}
\end{eqnarray}
Here, $\dx:=\vv{\th}-\vv{\th}^{(i)}$, $\bar{\ell}:=\ell\ap{\vv{\th}^{(i)}}$;
$\Gv:=\frac{\partial\ell}{\partial\vv{\th}}\ap{\vv{\th}^{(i)}}$ is
the gradient and $\Hm:=\pderi[\vv{\th}][2]{\ell}\ap{\vv{\th}^{(i)}}$
the Hessian matrix of $\ell$ at $\vv{\th}^{(i)}$. Analogously to
notation used above, we split $\dx$ into its first entry $\dxn_{0}$
and the remainder $\widetilde{\dx}$, $\Gv$ into $\G_{0}$ and $\widetilde{\Gv}$,
and write $\Hv_{0}$ for the first column of $\Hm$, $\widetilde{\Hm}$
for $\Hm$ without its first column and row, and split $\Hv_{0}$
into $\H_{00}$ and $\widetilde{\Hv}_{0}$.

In each iteration, VM seeks $\dxn_{0}^{*}$ and $\widetilde{\dx^{*}}$
that satisfy conditions \ref{enu:cond-val} and \ref{enu:cond-grad}.
Applying condition \ref{enu:cond-grad} to the approximation $\hat{\ell}^{\dxn}$
(equation (\ref{eq:approx-l})) yields
\begin{eqnarray}
\widetilde{\dx^{*}} & = & -\widetilde{\Hm}^{-1}\left(\widetilde{\Hv}_{0}\dxn_{0}+\widetilde{\Gv}\right).\label{eq:nuisance_diff}
\end{eqnarray}
Inserting (\ref{eq:approx-l}) and (\ref{eq:nuisance_diff}) into
condition \ref{enu:cond-val} gives us

\begin{eqnarray}
\ell^{*}\label{eq:f_diff} & = &
\frac{1}{2}\left(\H_{00}-\widetilde{\Hv}_{0}^{\top}\widetilde{\Hm}^{-1}\widetilde{\Hv}_{0}\right)\dxn_{0}^{*2} \nonumber\\
& & +\left(\G_{0}-\widetilde{\Gv}^{\top}\widetilde{\Hm}^{-1}\widetilde{\Hv}_{0}\right)\dxn_{0}^{*}+\bar{\ell}-\frac{1}{2}\widetilde{\Gv}^{\top}\widetilde{\Hm}^{-1}\widetilde{\Gv},
\end{eqnarray}
which can be solved for $\dxn_{0}^{*}$ if $\Hm$ is negative definite.
If equation (\ref{eq:f_diff}) has multiple solutions, \citet{venzon_method_1988}
choose the one that minimizes $\dx$ according to some norm. Our algorithm
RVM applies a different procedure and chooses the root that minimizes
the distance to $\th_{0}^{\max}$ without stepping into a region in
which the approximation (\ref{eq:approx-l}) is inaccurate. In section
\ref{subsec:Step-choice}, we provide further details and discuss
the case in which equation (\ref{eq:f_diff}) has no real solutions.

After each iteration, $\vv{\th}$ is updated according to the above
results:
\begin{eqnarray}
\vv{\th}^{(i+1)} & = & \vv{\th}^{(i)}+\dx^{*}.\label{eq:update}
\end{eqnarray}
If $\ell\ap{\vv{\th}^{(i+1)}}\approx\ell^{*}$ and $\pderi[\tilde{\vv{\th}}]{\ell}\ap{\vv{\th}^{(i+1)}}\approx0$
up to the desired precision, the search is terminated and $\vv{\th}^{(i+1)}$
is returned.

\begin{figure}
\begin{centering}
\includegraphics[scale=0.8]{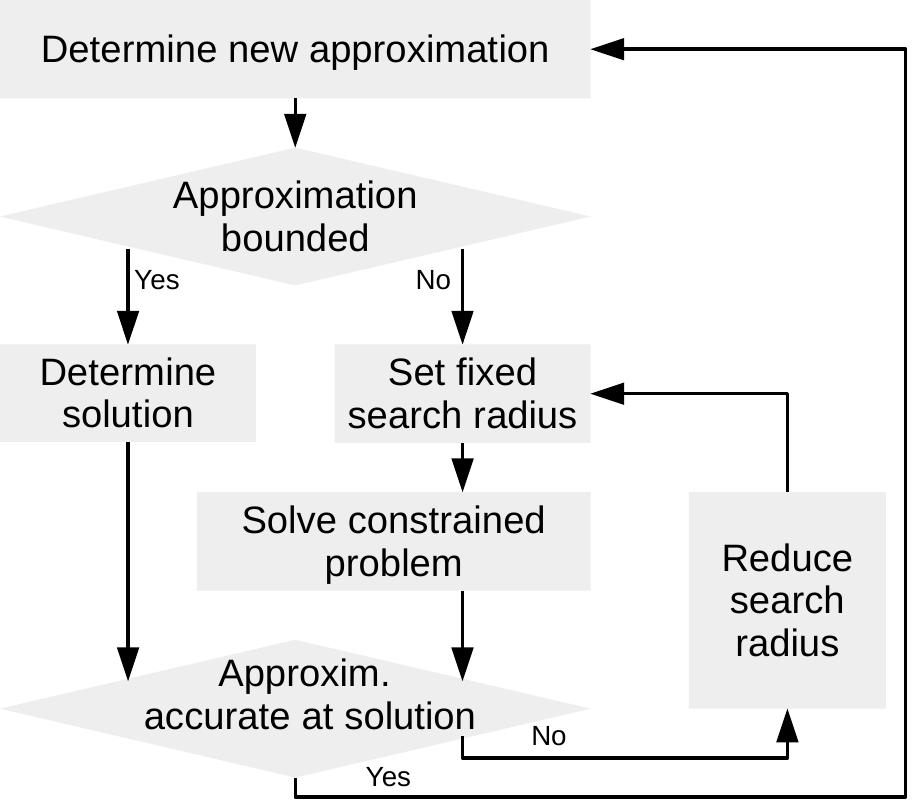}
\par\end{centering}
\caption[Flow chart for RVM]{Flow chart for RVM. The procedure is repeated until the termination
criterion is met and the result is returned.\label{fig:Flow-chart-RVM}}
\end{figure}
 The need to extend the original algorithm VM outlined above comes
from the following issues: (1) The quadratic approximation $\hat{\ell}$
may be imprecise far from the approximation point. In extreme cases,
updating $\vv{\th}$ as suggested could take us farther away from
the target $\vv{\th}^{*}$ rather than closer to it. (2) The approximation
$\hat{\ell}$ may be constant in some directions or be not bounded
above. In these cases, we may not be able to identify unique solutions
for $\dxn_{0}$ and $\widetilde{\dx}$, and the gradient criterion
in condition \ref{enu:cond-grad} may not characterize a maximum but
a saddle point or a minimum. (3) The limited precision of numerical
operations can result in discontinuities corrupting the results of
VM and hinder the algorithm from terminating.

To circumvent these problems, we introduce a number of extensions
to VM. First, we address the limited precision of the Taylor approximation
$\hat{\ell}$ with a trust region approach \citep{conn_trust-region_2000}.
That is, we constrain our search for $\dx^{*}$ to a region in which
the approximation $\hat{\ell}$ is sufficiently accurate. Second,
we choose some parameters freely if $\hat{\ell}$ is constant in some
directions and solve constrained maximization problems if $\hat{\ell}$
is not bounded above. In particular, we detect cases in which $\pl$
approaches an asymptote above $\ell^{*}$, which means that $\th_{0}$
is not estimable. Lastly, we introduce a method to identify and jump
over discontinuities as appropriate. An overview of the algorithm
is depicted as flow chart in Figure \ref{fig:Flow-chart-RVM}. Below,
we describe each of our extensions in detail.

\subsection{The trust region\label{subsec:The-trust-region}}

In practice, the quadratic approximation (\ref{eq:approx-l}) may
not be good enough to reach a point close to $\vv{\th}^{*}$ within
one step. In fact, since $\ell$ may be very ``non-quadratic'',
we might obtain a parameter vector for which $\ell$ and $\pderi[\tilde{\vv{\th}}]{\ell}$
are farther from $\ell^{*}$ and $\vv 0$ than in the previous iteration.
Therefore, we accept changes in $\th$ only if the approximation is
sufficiently accurate in the new point.

In each iteration $i$, we compute the new parameter vector, compare
the values of $\hat{\ell}$ and $\ell$ at the obtained point $\vv{\th}^{(i)}+\dx^{*}$,
and accept the step if, and only if, $\hat{\ell}$ and $\ell$ are
close together with respect to a given distance measure. If $\bar{\ell}$
is near the target $\ell^{*}$, we may also check the precision of
the gradient approximation $\pderi[\tilde{\vv{\th}}]{\hat{\ell}}$
to enforce timely convergence of the algorithm.

If we reject a step, we decrease the value $\dxn_{0}^{*}$ obtained
before, reduce the maximal admissible length $r$ of the nuisance
parameter vector and  solve the constrained maximization problem
\begin{eqnarray}
\widetilde{\dx^{*}} & = & \maxo{\dxt\,:\,\left|\dxt\right|\leq r}\hat{\ell}^{\dxn}\left(\dxn_{0},\dxt\right).\label{eq:constr-maximization}
\end{eqnarray}
As the quadratic subproblem (\ref{eq:constr-maximization}) appears
in classical trust-region algorithms, efficient solvers are available
\citep{conn_trust-region_2000}  and implemented in optimization
software, such as in the Python package Scipy \citep{jones_scipy:_2001}.

We check the accuracy of the approximation at the resulting point
$\vv{\th}^{(i)}+\dx^{*}$, decrease the search radius if necessary,
and continue with this procedure until the approximation is sufficiently
precise. The metric and the tolerance applied to measure the approximation's
precision may depend on how far the current log-likelihood $\bar{\ell}$
is from the target $\ell^{*}$. We suggest suitable precision measures
in section \ref{subsec:Parameters-and-distance-measures}.

Since it is often computationally expensive to compute the Hessian
$\Hm$, we desire to take as large steps $\dxn_{0}$ as possible.
However, it is also inefficient to adjust the search radius very often
to find the maximal admissible $\dxn_{0}^{*}$. Therefore, RVM first
attempts to make the unconstrained step given by equation (\ref{eq:approx-l}).
If this step is rejected, RVM determines the search radius with a
log-scale binary search between the radius of the unconstrained step
and the search radius accepted in the previous iteration. If even
the latter radius does not lead to a sufficiently precise result,
we update $\dxn_{0}^{*}$ and $r$ by factors $\beta_{0},\beta_{1}\in\left(0,1\right)$
so that $\dxn_{0}^{*}\leftarrow\beta_{0}\dxn_{0}^{*}$ and $r\leftarrow\beta_{1}r$.

\subsection{Linearly dependent parameters\label{subsec:Eliminating-linear-independencies}}

The right hand side of equation (\ref{eq:nuisance_diff}) is defined
only if the nuisance Hessian $\widetilde{\Hm}$ is invertible. If
$\widetilde{\Hm}$ is singular, the maximum with respect to the nuisance
parameters is not uniquely defined or does not exist at all. We will
consider the second case in the next section and focus on the first
case here.

There are multiple options to compute a psudo-inverse of a singular
matrix to solve underspecified linear equation systems \citep{rao_calculus_1967}.
A commonly used approach is the Moore-Penrose inverse \citep{penrose_generalized_1955},
which yields a solution with minimal norm \citep{rao_calculus_1967}.
This is a desirable property for our purposes, as the quadratic approximation
is generally most precise close to the approximation point. The Moore-Penrose
inverse can be computed efficiently with singular value decompositions
\citep{golub_calculating_1965}, which have also been applied to determine
the number of identifiable parameters in a model \citep{eubank_singular-value_1985,viallefont_parameter_1998}.

Whether or not a matrix is singular is often difficult to know precisely
due to numerical inaccuracies. The Moore-Penrose inverse is therefore
highly sensitive to a threshold parameter determining when the considered
matrix is deemed singular. As the Hessian matrix is typically computed
with numerical methods subject to error, it is often beneficial to
choose a high value for this threshold parameter to increase the robustness
of the method. Too large threshold values, however, can slow down
or even hinder convergence of the algorithm.

An alternative method to account for singular Hessian matrices is
to hold linearly dependent parameters constant until the remaining
parameters form a non-singular system. In tests, this approach appeared
to be more robust than applying the Moore-Penrose inverse. Therefore,
we used this method in our implementation. We provide details on this
method as well as test results in Supplementary Appendix A. Note that
we write $\widetilde{\Hm}^{-1}$ for this generalized inverse below.

To determine whether the approximate system has any solution when
$\widetilde{\Hm}$ is singular, we test whether $\wt{\dx^{*}}$ computed
according to equations (\ref{eq:nuisance_diff}) and (\ref{eq:f_diff})
indeed satisfies the necessary conditions for a maximum in the nuisance
parameters. That is, we check whether
\begin{equation}
0\la{\approx}\pder[\dxt]{}\hat{\ell}^{\dx}\laq\widetilde{\Hm}\wt{\dx^{*}}+\widetilde{\Hv}_{0}\dxn_{0}^{*}+\widetilde{\Gv}\label{eq:approximate-jac-test}
\end{equation}
holds up to a certain tolerance. If this is not the case, $\hat{\ell}$
is unbounded, and we proceed as outlined in the next section.

\subsection{Solving unbounded subproblems\label{subsec:Solving-unbouned-sub-problems}}

In each iteration, we seek the nuisance parameters $\tilde{\th}$
that maximize $\ell$ for the computed value of $\th_{0}$. Since
the log-likelihood function $\ell$ is bounded above, such a maximum
must exist in theory. However, the \emph{approximate }log-likelihood
$\hat{\ell}$ could be unbounded at times, which would imply that
the approximation is imprecise for large steps. Since we cannot identify
a global maximum of $\hat{\ell}$ if it is unbounded, we instead seek
the point maximizing $\hat{\ell}$ in the range where $\hat{\ell}$
is sufficiently accurate.

To test whether $\hat{\ell}$ is unbounded in the nuisance parameters,
we first check whether $\widetilde{\Hm}$ is negative semi-definite.
If $\widetilde{\Hm}$ is invertible, this test can be conducted by
applying a Cholesky decomposition on $-\widetilde{\Hm}$, which succeeds
if and only if $\widetilde{\Hm}$ is negative definite. If $\widetilde{\Hm}$
is singular, we use an eigenvalue decomposition. If all eigenvalues
are below a small threshold, $\widetilde{\Hm}$ is negative semi-definite.
To confirm that $\hat{\ell}$ is bounded, we also test whether equation
(\ref{eq:approximate-jac-test}) holds approximately if $\widetilde{\Hm}$
is singular (see section \ref{subsec:Eliminating-linear-independencies}).

If either of these tests fails, $\hat{\ell}$ is unbounded. In this
case, we set $\dxn_{0}^{*}\leftarrow r_{0}$, $r\leftarrow r_{1}$,
for some parameters $r_{0},r_{1}>0$ and solve the maximization problem
(\ref{eq:constr-maximization}). The parameters $r_{0}$ and $r_{1}$
can be adjusted and saved for future iterations to efficiently identify
the maximal admissible step. That is, we may increase (or reduce)
$\dxn_{0}^{*}$ and $r$ as long as (or until) $\hat{\ell}$ is sufficiently
precise. Thereby, we adjust the ratio of $\dxn_{0}^{*}$ and $r$
so that the likelihood increases: $\hat{\ell}^{\dxn}\ap{\dxn_{0}^{*},\wt{\dx^{*}}}>\bar{\ell}$.

\subsection{Step choice for the parameter of interest\label{subsec:Step-choice}}

Whenever $\hat{\ell}$ has a unique maximum in the nuisance parameters,
we compute $\dxn_{0}^{*}$ by solving equation (\ref{eq:f_diff}).
This equation can have one, two, or no roots. To discuss how $\dxn_{0}^{*}$
should be chosen in either of these cases, we introduce some helpful
notation. First, we write $\hat{\ell}_{\mt{PL}}\ap{\th_{0}}:=\maxo{\tilde{\th}}\hat{\ell}\ap{\th_{0},\tilde{\th}}$
for the profile log-likelihood function of the quadratic approximation.
Furthermore, we write in accordance with previous notation 
\begin{equation}
\hat{\ell}_{\mt{PL}}^{\dxn}\ap{\dxn_{0}}\la{:=}\hat{\ell}_{\mt{PL}}\ap{\th_{0}^{(i)}+\dxn_{0}}\laq a\dxn_{0}^{2}+p\dxn_{0}+q+\ell^{*}\label{eq:l^_PL}
\end{equation}
with $a:=\frac{1}{2}\left(\H_{00}-\widetilde{\Hv}_{0}\widetilde{\Hm}^{-1}\widetilde{\Hv}_{0}\right)$,
$p:=\G_{0}-\widetilde{\G}^{\top}\widetilde{\Hm}^{-1}\widetilde{\Hv}_{0}$,
and $q:=\bar{\ell}-\frac{1}{2}\widetilde{\G}^{\top}\widetilde{\Hm}^{-1}\widetilde{\G}-\ell^{*}$
(see equation (\ref{eq:f_diff})). 

Our choices of $\dxn_{0}^{*}$ attempt to increase $\th_{0}$ as
much as possible while staying in a region in which the approximation
$\hat{\ell}$ is reasonably accurate. The specific step choice depends
on the slope of the profile likelihood $\hat{\ell}_{\mt{PL}}^{\dxn}$
and on whether we have already exceeded $\th_{0}^{\max}$ according
to our approximation, i.e. $\hat{\ell}_{\mt{PL}}^{\dxn}\ap 0<\ell^{*}$.
Below, we will first assume that $\hat{\ell}_{\mt{PL}}^{\dxn}\ap 0>\ell^{*}$
and discuss the opposite case later.

\subsubsection{Case 1: decreasing profile likelihood}

\begin{figure*}[t]
\captionsetup[subfigure]{position=top,justification=raggedleft,margin=0pt, singlelinecheck=off}

\subfloat[\label{fig:PL-example-2roots}]{\includegraphics[width=0.3\textwidth]{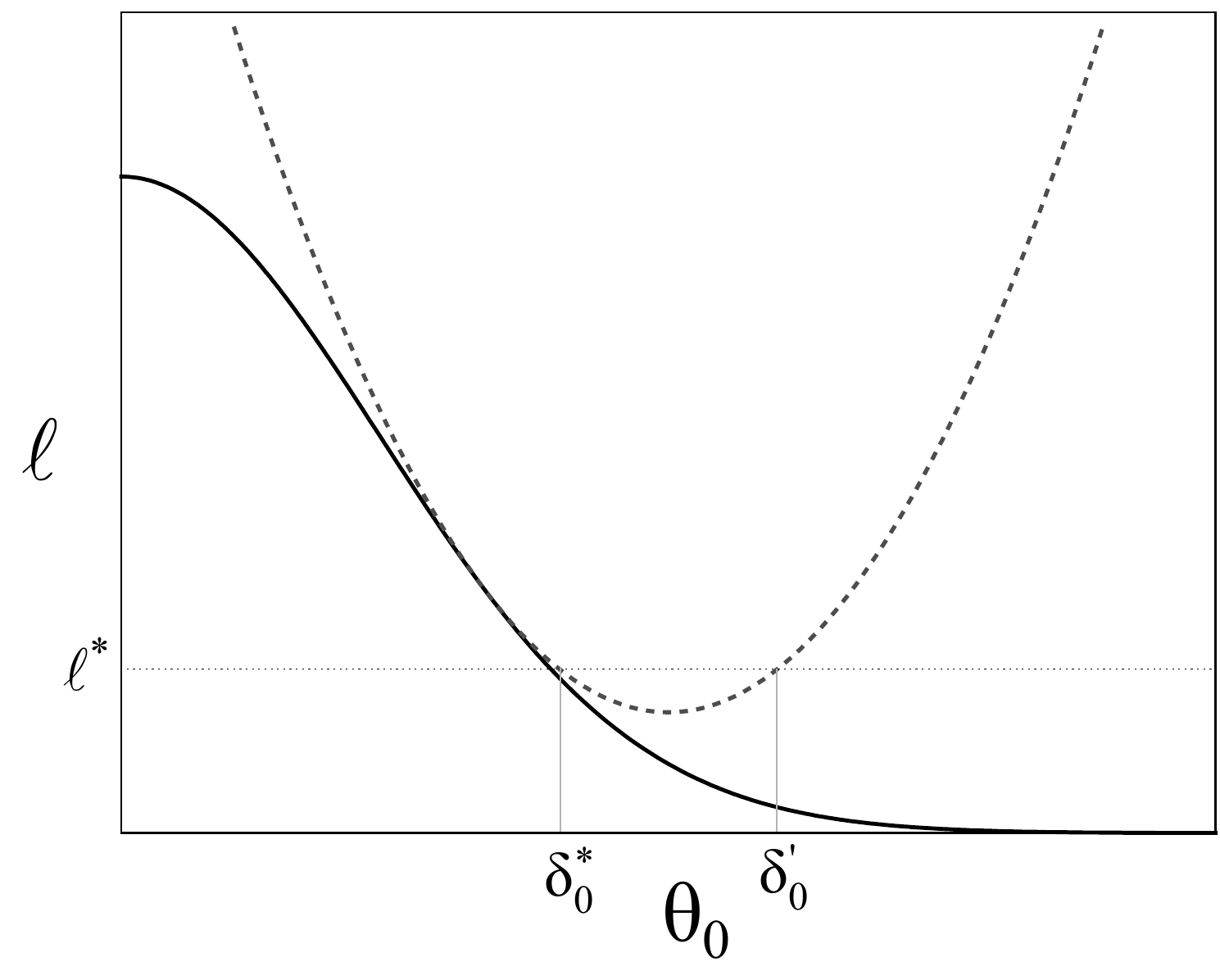}}\hspace*{\fill}\subfloat[\foreignlanguage{british}{\label{fig:Approx-before}}]{\includegraphics[width=0.3\textwidth]{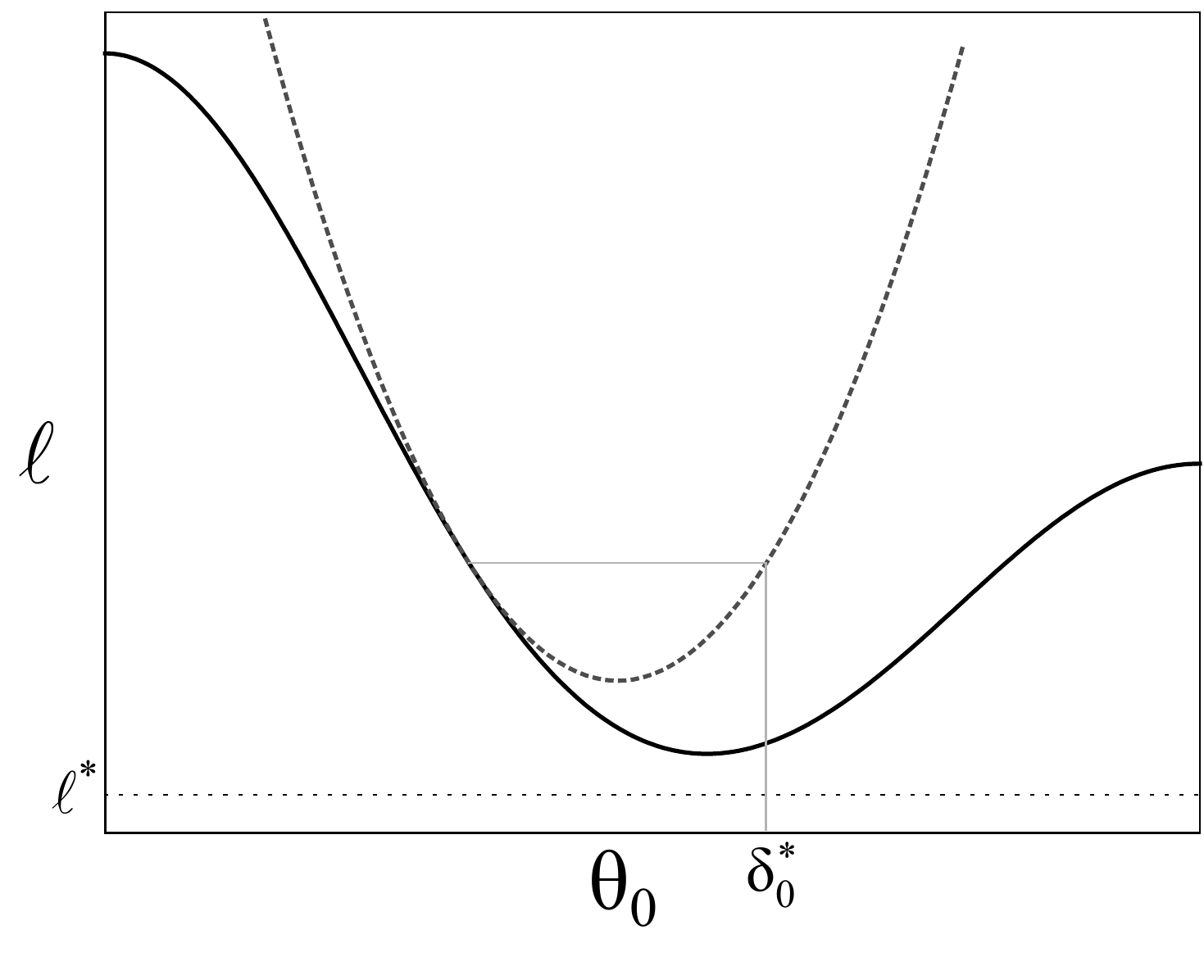}

}\hspace*{\fill}\subfloat[\foreignlanguage{british}{\label{fig:Approx-after}}]{\includegraphics[width=0.3\textwidth]{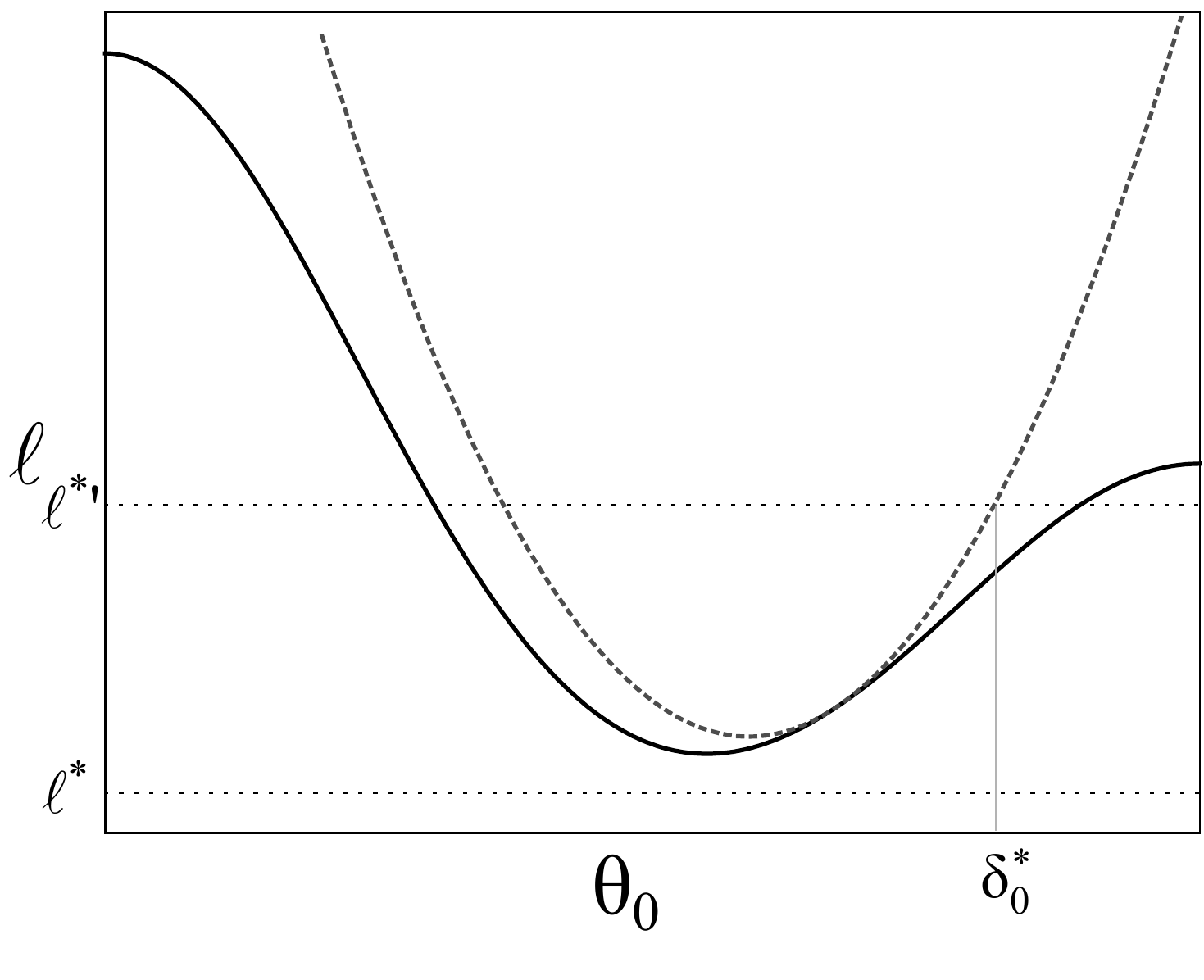}

}

\caption[Step choice for the parameter of interest in special cases]{Step choice for $\protect\th_{0}$ in special cases. The figures
depict the profile likelihood function $\ell_{\protect\mt{PL}}$ (solid
black), quadratic approximation $\hat{\ell}_{\protect\mt{PL}}$ (dashed
parabola), and the threshold log-likelihood $\ell^{*}$. (a) The approximation
has two roots $\delta_{0}^{*}$ and $\delta_{0}'$. Though the largest
root of $\ell$ is searched, the smaller root of $\hat{\ell}$ is
closest to the desired result. In fact, consistently choosing the
larger root would let the algorithm diverge. (b) If $\protect\pl$
is decreasing but $\hat{\ell}_{\protect\mt{PL}}$ does not assume
the threshold value $\ell^{*}$, we ``jump'' over the local minimum.
(c) If $\protect\pl$ is increasing but $\hat{\ell}_{\protect\mt{PL}}$
does not assume the threshold value $\ell^{*}$, we reset the target
value to an increased value $\ell^{*\prime}$.}
\end{figure*}

If the profile likelihood decreases at the approximation point, i.e.
$p<0$, we select the smallest positive root:
\begin{eqnarray}
\dxn_{0}^{*} & = & \begin{cases}
-\frac{q}{p} & \text{if }a=0\\
-\frac{1}{2a}\left(p+\sqrt{p^{2}-4aq}\right) & \text{else.}
\end{cases}
\end{eqnarray}
Choosing $\dxn_{0}^{*}>0$ ensures that the distance to the end point
$\th_{0}^{\max}$ decreases in this iteration. Choosing the smaller
positive root increases our trust in the accuracy of the approximation
and prevents potential convergence issues (see Figure \ref{fig:PL-example-2roots}).

If $\hat{\ell}_{\mt{PL}}^{\dxn}$ has a local minimum above the threshold
$\ell^{*}$, equation (\ref{eq:l^_PL}) does not have a solution,
and we may attempt to decrease the distance between $\hat{\ell}_{\mt{PL}}^{\dxn}$
and $\ell^{*}$ instead. This procedure, however, may let RVM converge
to a local minimum in $\hat{\ell}_{\mt{PL}}^{\dxn}$ rather than to
a point with $\hat{\ell}_{\mt{PL}}^{\dxn}=\ell^{*}$. Therefore, we
``jump'' over the extreme point by doubling the value of $\dxn_{0}^{*}$.
That is, we choose

\begin{eqnarray}
\dxn_{0}^{*} & = & -\frac{p}{a}
\end{eqnarray}
if $p^{2}<4aq$ (see Figure \ref{fig:Approx-before}).

\subsubsection{Case 2: increasing profile likelihood}

If the profile likelihood increases at the approximation point, i.e.
$p>0$, equation (\ref{eq:l^_PL}) has a positive root if and only
if $\hat{\ell}_{\mt{PL}}$ is concave down; $a<0$. We choose this
root whenever it exists:
\begin{eqnarray}
\dxn_{0}^{*} & = & -\frac{1}{2a}\left(p+\sqrt{p^{2}-4aq}\right).
\end{eqnarray}
However, if $\hat{\ell}_{\mt{PL}}$ grows unboundedly, equation (\ref{eq:l^_PL})
does not have a positive root. In this case, we change the threshold
value $\ell^{*}$ temporarily to a value $\ell^{*\prime}$ chosen
so that equation (\ref{eq:l^_PL}) has a solution with the updated
threshold (see Figure \ref{fig:Approx-after}). For example, we may
set 
\begin{eqnarray*}
\ell^{*\prime} & := & \max\left\{ \hat{\ell}_{\mt{PL}}^{\dxn}\ap 0+1,\frac{\bar{\ell}+\ell\ap{\hat{\vv{\th}}}}{2}\right\} .
\end{eqnarray*}
This choice ensures that a solution exists while at the same time
reaching local likelihood maxima quickly. After resetting the threshold,
we proceed as usual.

To memorize that we changed the threshold value $\ell^{*}$, we set
a flag $\code{maximizing:=True}$. In future iterations $j>i$, we
set the threshold $\ell^{*}$ back to its initial value $\ell_{0}^{*}$
and $\code{maximizing:=False}$ as soon as $\ell\ap{\vv{\th}^{(j)}}<\ell_{0}^{*}$
or $\hat{\ell}_{\mt{PL}}$ is concave down at the approximation point
$\vv{\th}^{(j)}$.

\subsubsection{Case 3: constant profile likelihood}

If the profile likelihood has a local extremum at the approximation
point, i.e. $p=0$, $a\neq0$, we proceed as in cases 1 and 2: if
$a>0$, we proceed as if $\hat{\ell}_{\mt{PL}}$ were increasing,
and if $a<0$, we proceed as if $\hat{\ell}_{\mt{PL}}$ were decreasing.
However, the approximate profile likelihood could also be constant,
$a=p=0$. In this case, we attempt to make a very large step to check
whether we can push $\th_{0}$ arbitrarily far. In section \ref{subsec:Identifying-non-estimable-parameters},
we discuss this procedure in greater detail.

\subsubsection{Profile likelihood below the threshold\label{subsec:Profile-likelihood-below-target}}

If the profile likelihood at the approximation point is below the
threshold, $\hat{\ell}_{\mt{PL}}^{\dxn}\ap 0<\ell^{*}$, we always
choose the smallest possible step: 
\begin{eqnarray}
\dxn_{0}^{*} & = & \begin{cases}
-\frac{1}{2a}\left(p+\sqrt{p^{2}-4aq}\right) & \text{if }a\neq0,\,p<0\\
-\frac{q}{p} & \text{if }a=0,\,p\neq0\\
-\frac{1}{2a}\left(p-\sqrt{p^{2}-4aq}\right) & \text{if }a\neq0,\,p>0.
\end{cases}\label{eq:d*_f<f*}
\end{eqnarray}
This shall bring us to the admissible parameter region as quickly
as possible.

As RVM rarely steps far beyond the admissible region in practice,
equation (\ref{eq:d*_f<f*}) usually suffices to define $\dxn_{0}^{*}$.
Nonetheless, if we find that $\hat{\ell}_{PL}^{\dx}$ has a local
maximum below the threshold, i.e. $p^{2}<4qa$, we may instead maximize
$\hat{\ell}_{PL}^{\dxn}$ as far as possible: 
\begin{eqnarray}
\dxn_{0}^{*} & = & -\frac{p}{2a}.
\end{eqnarray}
If we have already reached a local maximum ($p\approx0$), we cannot
make a sensible choice for $\dxn_{0}$. In this case, we may recall
the iteration $k:=\argmaxo{j\,:\,\ell(\th^{(j)})\geq\ell^{*}}\th_{0}^{(j)}$,
in which the largest admissible $\th_{0}$ value with $\ell\ap{\th^{(k)}}\geq\ell^{*}$
has been found so far, and conduct a binary search between $\th^{(i)}$
and $\th^{(k)}$ until we find a point $\th^{(i+1)}$ with $\ell\ap{\th^{(i+1)}}\geq\ell^{*}$.

\subsection{Identifying inestimable parameters\label{subsec:Identifying-non-estimable-parameters}}

In some practical scenarios, the profile log-likelihood $\ell_{\mt{PL}}$
will never fall below the threshold $\ell^{*}$, which means that
the considered parameter is not estimable. In these cases, RVM may
not converge. However, often it is possible to identify inestimable
parameters by introducing a step size limit $\dxn_{0}^{\max}$. If
the computed step exceeds the maximal step size, $\dxn_{0}^{*}>\dxn_{0}^{\max}$
and the current function value exceeds the threshold value, i.e. $\bar{\ell}\geq\ell^{*}$,
we set $\dxn_{0}^{*}:=\dxn_{0}^{\max}$ and compute the corresponding
nuisance parameters. If the resulting log-likelihood $\ell\ap{\th^{(i)}+\dx^{*}}$
is not below the threshold $\ell^{*}$, we let the algorithm terminate,
raising a warning that the parameter $\th_{0}$ is not estimable.
If $\ell\ap{\th^{(i)}+\dx^{*}}<\ell^{*}$, however, we cannot draw
this conclusion and decrease the step size until the approximation
is sufficiently close to the original function.

The criterion suggested above may not always suffice to identify inestimable
parameters. For example, if the profile likelihood is constant but
the nuisance parameters maximizing the likelihood change non-linearly,
RVM may not halt. For this reason, and also to prevent unexpected
convergence issues, it is advisable to introduce an iteration limit
to the algorithm. If the iteration limit is exceeded, potential estimability
issues issues may be investigated further.

\subsection{Discontinuities\label{subsec:Handling-discontinuities}}

RVM is based on quadratic approximations and requires therefore that
$\ell$ is differentiable twice. Nonetheless, discontinuities can
occur due to numerical imprecision even if the likelihood function
is continuous in theory. Though we may still be able to compute the
gradient $\G$ and the Hessian $\Hm$ in these cases, the resulting
quadratic approximation will be inaccurate even if we take very small
steps. Therefore, these discontinuities could hinder the algorithm
from terminating.

To identify discontinuities, we define a minimal step size $\epsilon$,
which may depend on the gradient $\Gv$. If we reject a step with
small length $\left|\dx^{*}\right|\le\epsilon$, we may conclude that
$\ell$ is discontinuous at the current approximation point $\vv{\th}^{(i)}$.
To determine the set $D$ of parameters responsible for the issue,
we decompose $\dx^{*}$ into its components. We initialize $D\leftarrow\emptyset$
and consider, with the $j^{\text{th}}$ unit vector $\vv e_{j}$,
the step $\dx^{*\prime}:=\sum_{j\leq k,\,j\neq D}\vv e_{j}\dxn_{j}^{*}$
until $\hat{\ell}^{\delta}\ap{\dx^{*\prime}}\not\approx\ell^{\delta}\ap{\dx^{*\prime}}$
for some $k<n$. When we identify such a component, we add it to the
set $D$ and continue the procedure.

If we find that $\ell$ is discontinuous in $\th_{0}$, we check whether
the current nuisance parameters maximize the likelihood, i.e. $\ell$
is bounded above and $\widetilde{\Gv}$ is approximately $\vv 0$.
If the nuisance parameters are not optimal, we hold $\th_{0}$ constant
and maximize $\ell$ with respect to the nuisance parameters. Otherwise,
we conclude that the profile likelihood function has a jump discontinuity.
In this case, our action depends on the current log-likelihood value
$\bar{\ell}$, the value of $\ell$ at the other end of the discontinuity,
and the threshold $\ell^{*}$.
\begin{itemize}
\item If $\ell\ap{\vv{\th}^{(i)}+\vv e_{0}\dxn_{0}^{*}}\geq\ell^{*}$ or
$\ell\ap{\vv{\th}^{(i)}}<\ell\ap{\vv{\th}^{(i)}+\vv e_{0}\dxn_{0}^{*}}$,
we accept the step regardless of the undesirably large error.
\item If $\ell\ap{\vv{\th}^{(i)}+\vv e_{0}\dxn_{0}^{*}}<\ell^{*}$ and $\ell\ap{\vv{\th}^{(i)}}\geq\ell^{*}$
, we terminate and return $\th_{0}^{(i)}$ as the bound of the confidence
interval.
\item Otherwise, we cannot make a sensible step and try to get back into
the admissible region by conducting the binary search procedure we
have described in section \ref{subsec:Profile-likelihood-below-target}.
\end{itemize}
If $\ell$ is discontinuous in variables other than $\th_{0}$, we
hold the variables constant whose change decreases the likelihood
and repeat the iteration with a reduced system. After a given number
of iterations, we release these parameters again, as $\th$ may have
left the point of discontinuity.

Since we may require that not only $\hat{\ell}$ but also its gradient
are well approximated, a robust implementation of RVM should also
handle potential gradient discontinuities. The nuisance parameters
causing the issues can be identified analogously to the procedure
outlined above. All components in which the gradient changes its sign
from positive to negative should be held constant, as the likelihood
appears to be in a local maximum in these components. The step in
the remaining components may be accepted regardless of the large error.

\subsection{Suitable parameters and distance measures\label{subsec:Parameters-and-distance-measures}}

The efficiency of RVM depends highly on the distance measures and
parameters applied when assessing the accuracy of the approximation
and updating the search radius of the constrained optimization problems.
If the precision measures are overly conservative, then many steps
will be needed to find $\vv{\th}^{*}$. If the precision measure is
too liberal, in turn, RVM may take detrimental steps and might not
even converge.

We suggest the following procedure: (1) we always accept forward steps
with $\dxn_{0}^{*}\geq0$ if the true likelihood is larger than the
approximate likelihood, $\ell^{\delta}\ap{\dx^{*}}\geq\hat{\ell}^{\delta}\ap{\dx^{*}}$.
(2) If the approximate likelihood function is unbounded, we require
that the likelihood increases $\ell^{\delta}\ap{\dx^{*}}\geq\bar{\ell}$.
This requirement helps RVM to return quickly to a region in which
the approximation is bounded. However, if the step size falls below
the threshold used to detect discontinuities, we may relax this constraint
so that less time must be spent to detect potential discontinuities.
(3) If we are outside the admissible region, i.e. $\bar{\ell}<\ell^{*}$,
we enforce that we get closer to the target likelihood: $\left|\ell^{\delta}\ap{\dx^{*}}-\ell^{*}\right|<\left|\bar{\ell}-\ell^{*}\right|$.
This reduces potential convergence issues. (4) We require that 
\begin{eqnarray}
\frac{\left|\hat{\ell}^{\delta}\ap{\dx^{*}}-\ell^{\delta}\ap{\dx^{*}}\right|}{\left|\bar{\ell}-\ell^{*}\right|} & \leq & \gamma
\end{eqnarray}
for a constant $\gamma$. That is, the required precision depends
on how close we are to the target. This facilitates fast convergence
of the algorithm. The constant $\gamma\in\left(0,1\right)$ controls
how strict the precision requirement is. In tests, $\gamma=\frac{1}{2}$
appeared to be a good choice. (5) If we are close to the target, $\ell^{\delta}\ap{\dx^{*}}\approx\ell^{*}$,
we also require that the gradient estimate is precise:
\begin{eqnarray}
\frac{\left|\pder[\tilde{\th}]{\hat{\ell}^{\delta}}\ap{\dx^{*}}-\pder[\tilde{\th}]{\ell^{\delta}}\ap{\dx^{*}}\right|}{\left|\Gv\right|} & \leq & \gamma.
\end{eqnarray}
This constraint helps us to get closer to a maximum in the nuisance
parameters. Here, we use the $\mathcal{L}_{2}$ norm.

When we reject a step because the approximation is not sufficiently
accurate, we adjust $\delta_{0}^{*}$ and solve the constrained maximization
problem (\ref{eq:constr-maximization}) requiring $\left|\dxt\right|\leq r$.
To ensure that the resulting step does not push the log-likelihood
below the target $\ell^{*}$, the radius $r$ should not be decreased
more strongly than $\delta_{0}^{*}$.  In tests, adjusting $r$ by
a factor $\beta_{1}:=1.5$ whenever $\delta_{0}^{*}$ is adjusted
by factor $\beta_{0}:=2$ appeared to be a good choice.

\subsection{Confidence intervals for functions of parameters}

Often, modelers are interested in confidence intervals for functions
$f\ap{\vv{\th}}$ of the parameters. A limitation of VM and VMR is
that such confidence intervals cannot be computed directly with these
algorithms. However, this problem can be solved approximately by considering
a slightly changed likelihood function. We aim to find 
\begin{eqnarray}
\phi^{\max} & = & \maxo{\vv{\th}\in\Theta\,:\,\ell\ap{\vv{\th}}\geq\ell^{*}}f\ap{\vv{\th}}\label{eq:constrained-max-f}
\end{eqnarray}
or the respective minimum. Define 
\begin{eqnarray}
\check{\ell}\ap{\phi,\vv{\th}} & := & \ell\ap{\vv{\th}}-\frac{1}{2}\left(\frac{f\ap{\vv{\th}}-\phi}{\varepsilon}\right)^{2}\chi_{1,1-\alpha}^{2},
\end{eqnarray}
with a small constant $\varepsilon$. Consider the altered maximization
problem 
\begin{eqnarray}
\check{\phi}^{\max} & = & \maxo{\vv{\th}\in\Theta\,:\,\check{\ell}\ap{\phi,\vv{\th}}\geq\ell^{*}}\phi,\label{eq:constrained-max-f-changed}
\end{eqnarray}
which can be solved with VM or RVM.

We argue that a solution to (\ref{eq:constrained-max-f-changed})
is an approximate solution to (\ref{eq:constrained-max-f}), whereby
the error is bounded by $\varepsilon$. Let $\ap{\phi^{\max},\vv{\th}^{*}}$
be a solution to problem (\ref{eq:constrained-max-f}) and $\ap{\check{\phi}^{\max},\check{\vv{\th}}^{*}}$
a solution to problem (\ref{eq:constrained-max-f-changed}). Since
$\phi^{\max}=f\ap{\vv{\th}^{*}}$, it is $\check{\ell}\ap{\phi^{\max},\vv{\th}^{*}}=\ell\ap{\vv{\th}^{*}}\geq\ell^{*}$.
Therefore, $\ap{\phi^{\max},\vv{\th}^{*}}$ is also a feasible solution
to (\ref{eq:constrained-max-f}), and it follows that $\check{\phi}^{\max}\geq\phi^{\max}$.
At the same time, $\check{\ell}\ap{\phi,\vv{\th}}\leq\ell\ap{\vv{\th}}$,
which implies that $f\ap{\check{\vv{\th}}^{*}}\leq f\ap{\vv{\th}^{*}}$,
since $\vv{\th}^{*}$ maximizes $f$ over a domain larger than the
feasibility domain of (\ref{eq:constrained-max-f-changed}). In conclusion,
$f\ap{\check{\vv{\th}}^{*}}\leq f\ap{\vv{\th}^{*}}=\phi^{\max}\leq\check{\phi}^{\max}$.
Lastly, 
\begin{eqnarray}
\ell^{*} & = & \ell\ap{\hat{\vv{\th}}}-\frac{1}{2}\chi_{1,1-\alpha}^{2}\la{\leq}\check{\ell}\ap{\check{\phi}^{\max},\check{\vv{\th}}^{*}} \nonumber \\
& = & \ell\ap{\check{\vv{\th}}^{*}}-\frac{1}{2}\left(\frac{f\ap{\check{\vv{\th}}^{*}}-\check{\phi}^{\max}}{\varepsilon}\right)^{2}\chi_{1,1-\alpha}^{2}.
\label{eq:error-bound-ineq}
\end{eqnarray}
Simplifying (\ref{eq:error-bound-ineq}) yields $\left|f\ap{\check{\vv{\th}}^{*}}-\check{\phi}^{\max}\right|\leq\varepsilon$.
Thus, $\left|\phi^{\max}-\check{\phi}^{\max}\right|\leq\varepsilon$.

Though it is possible to bound the error by an arbitrarily small constant
$\varepsilon$ in theory, care must be taken if the function $f\ap{\vv{\th}}$
is not well-behaved, i.e. strongly nonlinear. In theses cases, overly
small values for $\varepsilon$ may slow down convergence.

Note that the suggested procedure may seem to resemble the approach
of \citet{neale_use_1997}, who also account for constraints by adding
the squared error to the target function. However, unlike \citet{neale_use_1997},
the approach suggested above bounds the error in the confidence interval
bound, not the error of the constraint. Furthermore, we do not square
the log-likelihood function, which would worsen nonlinearities and
could thus make optimization difficult. Therefore, our approach is
less error-prone than the method by \citet{neale_use_1997}.

\section{Tests\label{sec:Tests}}

To compare the presented algorithm to existing methods, we applied
RVM, the classic VM, and five other algorithms to benchmark problems
and compared the robustness and performance of the approaches. Below
we review the implemented methods. Then we introduce the benchmark
problems, before we finally present the benchmark results.

\subsection{Methods implemented for comparison\label{subsec:Method}}

Besides RVM and VM, we implemented three methods that repeatedly evaluate
the profile likelihood function and two methods that search for the
confidence intervals directly. We implemented all methods in the programming
language Python version 3.7 and made use of different optimization
routines implemented or wrapped in the scientific computing library
Scipy \citep{jones_scipy:_2001}.

First, we implemented a grid search for the confidence bounds. The
approach uses repeated Lagrangian constrained optimizations and may
resemble the method by \citet{diciccio_implementation_1991}; however,
rather than implementing the algorithm by \citet{diciccio_implementation_1991},
we applied the constrained optimization algorithm by \citet{lalee_implementation_1998},
which is a trust-region approach and may thus be more robust than
the method by \citet{diciccio_implementation_1991}. Furthermore,
the algorithm by \citet{lalee_implementation_1998} was readily implemented
in Scipy.

We conducted the grid search with a naive step size of $0.2$, which
we repeatedly reduced by factor $2$ close to the threshold log-likelihood
$\ell^{*}$ until the desired precision was achieved. To account for
unidentifiable parameters, we attempted one large step ($1000$ units)
if the algorithm did not terminate in the given iteration limit. We
considered a parameter as unbounded if this step yielded a log-likelihood
above the target value $\ell^{*}$.

Second, we implemented a quadratic bisection method for root finding
on $\pl$ (cf. \citealp{ren_algorithm_2019}). Initially we chose
a step size of $1$. Afterwards, we computed the step of $\th_{0}$
based on a quadratic interpolation between the MLE $\hat{\th}_{0}$,
the maximal value of $\th_{0}$ for which we found $\pl\ap{\th_{0}}>\ell^{*}$
and the smallest identified value of $\th_{0}$ with $\pl\ap{\th_{0}}<\ell^{*}$.
Until a point $\th_{0}$ with $\pl\ap{\th_{0}}<\ell^{*}$ was identified,
we interpolated $\pl$ between $\hat{\th}_{0}$ and the two largest
evaluated values $\th_{0}$. When only two points were available or
the approximation of $\pl$ did not assume the target value, we introduced
the additional constraint $\deri[\th_{0}][][\pl]=0$. Using a quadratic
rather than a linear interpolation for bisection has the advantage
that the algorithm converges faster if the profile log-likelihood
function is convex or quadratic. To evaluate $\pl$, we applied sequential
least squares programming \citep{kraft_software_1988}, which is the
default method for constrained optimization in Scipy.

Third, we implemented a binary search with an initial step of $1$.
Until a value $\th_{0}$ with $\pl\ap{\th_{0}}<\ell^{*}$ was found,
we increased $\th_{0}$ by factor $10$. This preserves the logarithmic
runtime of the algorithm if the problem has a solution. To broaden
the range of tested internal optimization routines, we used a different
method to evaluate $\pl$ than in the bisection method: we fixed $\th_{0}$
at the desired value and performed an unconstrained optimization on
the nuisance parameters. Here, we used the quasi-Newton method by
Broyden, Fletcher, Goldfarb, and Shanno (BFGS; see \citealp{nocedal_numerical_2006},
pp. 136).

To test methods that search for the confidence interval end points
directly, we solved problem (\ref{eq:constrained-max}) with sequential
least squares programming \citep{kraft_software_1988}. Furthermore,
we implemented the approximate method by \citet{neale_use_1997}.
They transform the constrained maximization problem (\ref{eq:constr-maximization})
to an unconstrained problem by considering the sum of the parameter
of interest $\th_{0}$ and the squared error between the target $\ell^{*}$
and the log-likelihood. Minimization of this target function yields
a point in which the target log-likelihood is reached approximately
and the parameter of interest is minimal. Again, we used the method
BFGS for minimization (see above).

Finally, we implemented Wald's method to assess the need to apply
any profile likelihood method.

\subsection{Benchmark problem}

To investigate the performances of the implemented methods, we applied
the algorithms to a benchmark problem with variable parameter number
and data set size. We considered a logistic regression problem with
$n$ count data covariates $c_{ij}$, $j\in\left\{ 1,\dots,n\right\} $
for each data point $i\in\left\{ 1,\dots,N\right\} $. We assumed
that the impact of the covariates levels off at high values and considered
therefore the transformed covariates $c_{ij}^{\alpha_{j}}$ with $\alpha\in\left(0,1\right)$.
This is not only reasonable in many real world problems but also makes
likelihood maximization a computationally challenging problem if not
enough data are available to achieve asymptotic normality of the MLE.
Hence, this scenario gives insights into the performance of the implemented
methods in challenging realistic problems. The benchmark model's probability
mass function for a data point $X_{i}$ was thus given by 
\begin{eqnarray}
\pr{X_{i}=1} & = & \left(1+\exp\ap{-\beta_{0}-\smo j\beta_{j}c_{ij}^{\alpha_{j}}}\right)^{-1}
\end{eqnarray}
and $\pr{X_{i}=0}=1-\pr{X_{i}=1}$.

We drew the covariate values randomly from a negative binomial distribution
with mean $5$ and variance $10$. The negative binomial distribution
is commonly used to model count data \citep{gardner_regression_1995}
and thus suited to represent count covariates. To simulate the common
case that covariates are correlated, we furthermore drew the value
for every other covariate from a binomial distribution with the respective
preceding covariate as count parameter. That is, for uneven $j$,
\begin{eqnarray*}
c_{i,j+1} & \sim & \mt{Binomial}\ap{c_{i,j},p},
\end{eqnarray*}
with $p=0.2$ in our simulations. To avoid numerical problems arising
when covariates with value $0$ are raised to the power $0$, we added
a small positive perturbation to the count values. That way, we achieved
that $0^{0}$ was defined to be $1$. We chose the parameters $\alpha_{j}$
and $\beta_{j}$ so that the data were balanced, i.e. the frequency
of $0$s and $1$s was approximately even. Refer to Supplementary
Appendix B for the parameter values we used.

\subsection{Test procedure}

\begin{figure*}
\captionsetup[subfigure]{position=top,justification=raggedleft,margin=0pt, singlelinecheck=off}

\hspace{-0.02\textwidth}\subfloat[\label{fig:res-500}]{\includegraphics[width=0.34\textwidth]{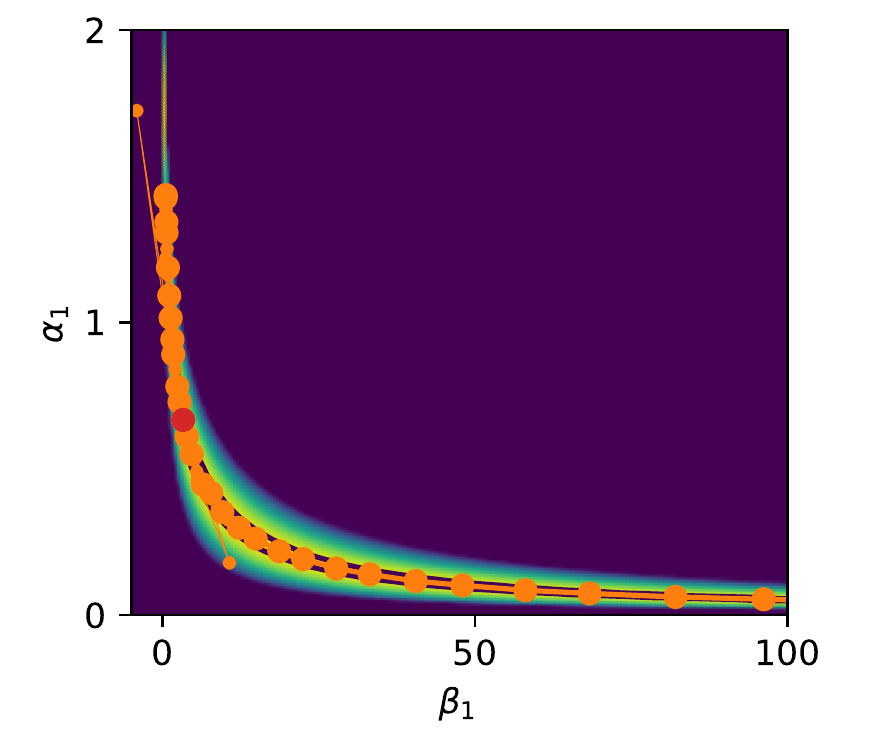}}\hspace*{\fill}\subfloat[\label{fig:Res-1000}]{\includegraphics[width=0.34\textwidth]{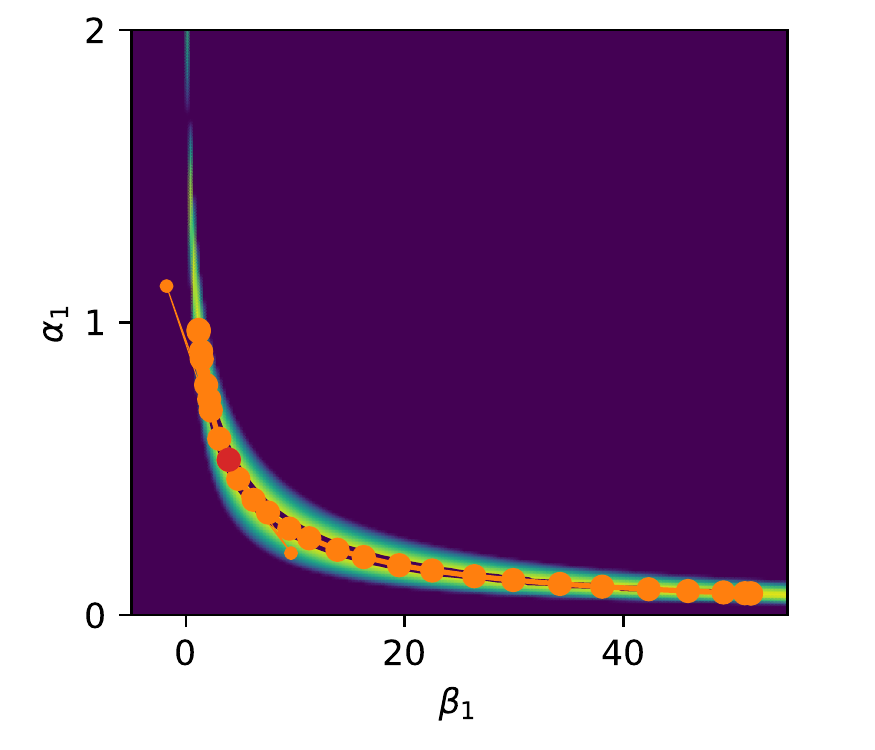}

}\hspace*{\fill}\subfloat[\label{fig:res-10000}]{\includegraphics[width=0.34\textwidth]{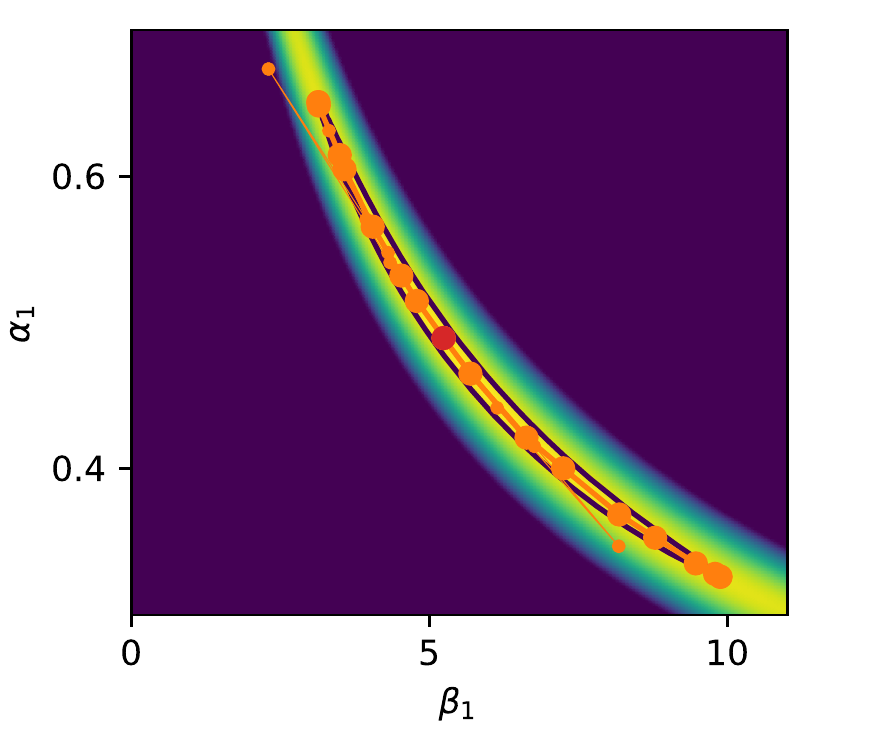}

}

\caption[Likelihood surface of the benchmark model with different data set
sizes]{Likelihood surface of the $3$-parameter benchmark model with different
data set sizes $N$. As $N$ increases, the confidence region becomes
smaller and closer to an elliptic shape. The orange dots depict the
accepted (large dots) and rejected (small dots) steps of RVM searching
for a confidence interval for $\beta_{1}$. RVM follows the ridge
of the likelihood surface. The red dot shows the location of the MLE
$\hat{\protect\th}$. The background color depicts the respective
maximal log-likelihood for the given $\alpha_{1}$ and $\beta_{1}$
ranging from $\protect\leq\hat{\ell}-50$ (dark blue) to $\hat{\ell}$
(yellow). The solid blue line denotes the target log-likelihood $\ell^{*}$
for a $95\%$ confidence interval. (a) $N=500$; (b) $N=1000$; (c)
$N=10000$. \label{fig:Likelihood-surface} }
\end{figure*}

To test the algorithms in a broad range of scenarios and assess how
their performance is impacted by model characteristics, we considered
a model with $1$ covariate ($3$ parameters), a model with $5$ covariates
($11$ parameters), and a generalized linear model (GLM) with $10$
covariates, in which the powers $\alpha_{j}$ were set to $1$ ($11$
parameters). Furthermore, we varied the sizes of the simulated data
sets, ranging between $N=500$ and $N=10000$ for the models with
transformed covariates and $N=50$ and $N=1000$ for the GLM. In Figure
\ref{fig:Likelihood-surface}, we depict the impact of $N$ on the
shape of the likelihood function and thus the difficulty of the problem.

For each considered set of parameters, we generated $200$ realizations
of covariates and training data from the model described in the previous
section. We determined the maximum likelihood estimator by maximizing
the log-likelihood with the method BFGS and refined the estimate with
an exact trust region optimizer \citep{conn_trust-region_2000}. Then,
we applied each of the implemented algorithms to each data set and
determined the algorithms' success rates and efficiencies.

As the likelihood functions of the tested models decrease drastically
at $\alpha_{j}=0$, potentially causing some algorithms to fail, we
constrained the $\alpha_{j}$ to non-negative values. Tho that end,
we considered transformed parameters $\alpha_{j}':=\ln\ap{\exp\ap{\alpha_{j}}-1}$.
Such transformations are reasonable whenever the parameter range is
naturally constrained from a modeling perspective. Nonetheless, we
evaluated the results of the tested algorithms based on the back-transformed
parameters $\alpha_{j}$.

We measured the algorithms' success based on their ability to solve
problem (\ref{eq:constrained-max}) rather than their capability to
determine the true confidence intervals for the parameters. Though
profile likelihood confidence intervals are usually highly accurate,
they rely on the limiting distribution of the likelihood ratio statistic.
Therefore, algorithms could fail to solve optimization problem (\ref{eq:constrained-max})
but, by coincidence, return a result close to the true confidence
interval bound and vice versa. To exclude such effects and circumvent
the high computational effort required to determine highly precise
confidence intervals with sampling methods, we determined the ``true''
confidence interval bound by choosing the widest confidence interval
bound obtained by either of the tested methods provided it was admissible,
i.e. $\ell\ap{\th^{\max}}\geq\ell^{*}$ up to a permissible error
of $0.001$.

We considered an algorithm successful if (1) the returned result was
within a $\pm5\%$ range of the true confidence interval bound or
had an error below $0.001$, and (2) the algorithm reported convergence.
That is, to be deemed successful, an algorithm had to both return
the correct result and also claim that it found the correct solution.
The latter constraint ensures that if none of the algorithms converges
successfully, even the one with the best result is not considered
successful.

As many of the tested methods rely on general optimizers without specific
routines to identify situations with divergent solutions, we considered
parameters with confidence interval bounds exceeding $\left[-1000,1000\right]$
in the transformed parameter space as unbounded. Consequently, all
algorithms returning a larger confidence interval were considered
successful.

We limited the runtime of all methods except the pre-implemented optimizers
by introducing a step limit of $200$. If convergence was not reached
within this number of steps, the algorithms were viewed unsuccessful
except for the case with inestimable parameters.

To test whether some methods tend return misleading results, we determined
the mean absolute error between the returned and the true confidence
interval bounds when algorithms reported success. As this quantity
can be dominated by outliers, we also determined the mean of all errors
below $10$ and the frequency of errors beyond $10$.

We measured the computational speed of the different methods by recording
the number of function evaluations required until termination. This
provides us with precise benchmark results independent of hardware
and implementation details. To display a potential trade-off between
robustness (success rate) and speed (number of function evaluations),
we did not consider cases in which convergence was not reached. That
way, internal stopping criteria did not affect the results.

The specific advantage of some optimization algorithms is in not requiring
knowledge of the Hessian matrix. As computing the Hessian is necessary
for RVM and may reduce the algorithm's performance compared to other
methods, we included the number of function evaluations required to
determine the Hessian and the gradient in the recorded count of function
evaluations. We computed gradients and Hessian matrices with a complex
step method \citep{lai_new_2005} implemented in the Python package
numdifftools \citep{brodtkorb_numdifftools_2019}.

\subsection{Results\label{subsec:Results}}

To get an impression of how RVM acts in practice, we plotted the trajectory
of RVM along with ancillary function evaluations in Figure \ref{fig:Likelihood-surface}.
It is visible that the algorithm stays on the ``ridge'' of the likelihood
surface even if the admissible region is strongly curved. This makes
RVM efficient.

\begin{figure*}
\begin{tabular}[t]{>{\raggedleft}p{0.5cm}>{\centering}p{0.27\textwidth}>{\centering}p{0.27\textwidth}>{\centering}p{0.27\textwidth}}
\begin{turn}{90}
\end{turn} & \centering{}\textsf{\footnotesize{}\hspace*{0.8cm}}%
\begin{minipage}[t]{3cm}%
\begin{center}
\textsf{\footnotesize{}A}
\par\end{center}%
\end{minipage} & \centering{}\textsf{\footnotesize{}\hspace*{0.8cm}}%
\begin{minipage}[t]{3cm}%
\begin{center}
\textsf{\footnotesize{}B}
\par\end{center}%
\end{minipage} & \centering{}\textsf{\footnotesize{}\hspace*{0.8cm}}%
\begin{minipage}[t]{3cm}%
\begin{center}
\textsf{\footnotesize{}C}
\par\end{center}%
\end{minipage}\tabularnewline
\begin{turn}{90}
\textsf{\footnotesize{}\hspace*{1cm}}%
\begin{minipage}[t]{3cm}%
\begin{center}
\textsf{\footnotesize{}Success rate}
\par\end{center}%
\end{minipage}
\end{turn} & \includegraphics[width=0.3\textwidth]{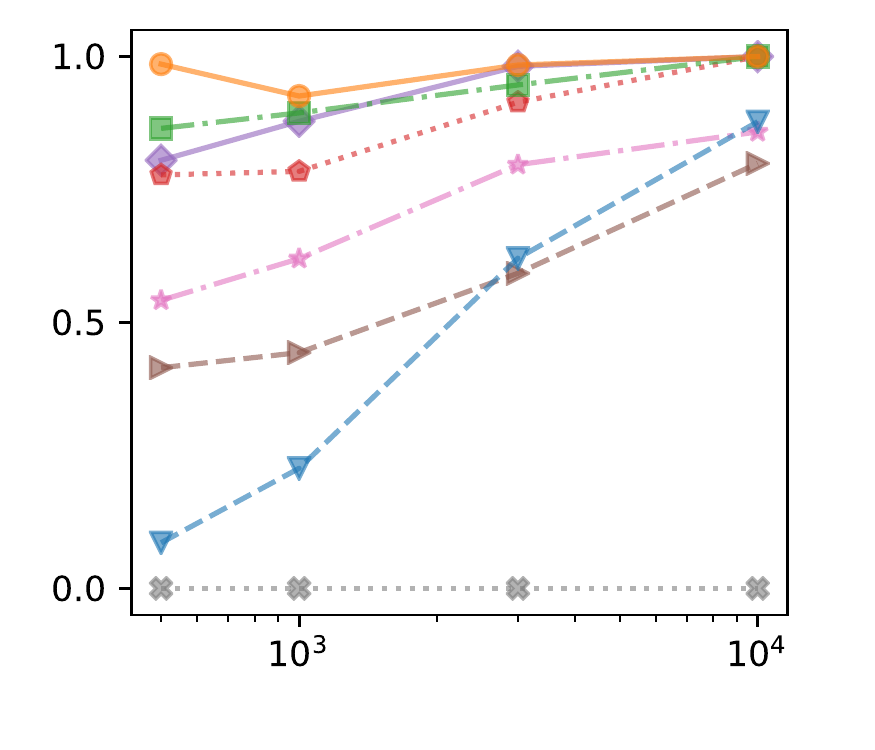} & \includegraphics[width=0.3\textwidth]{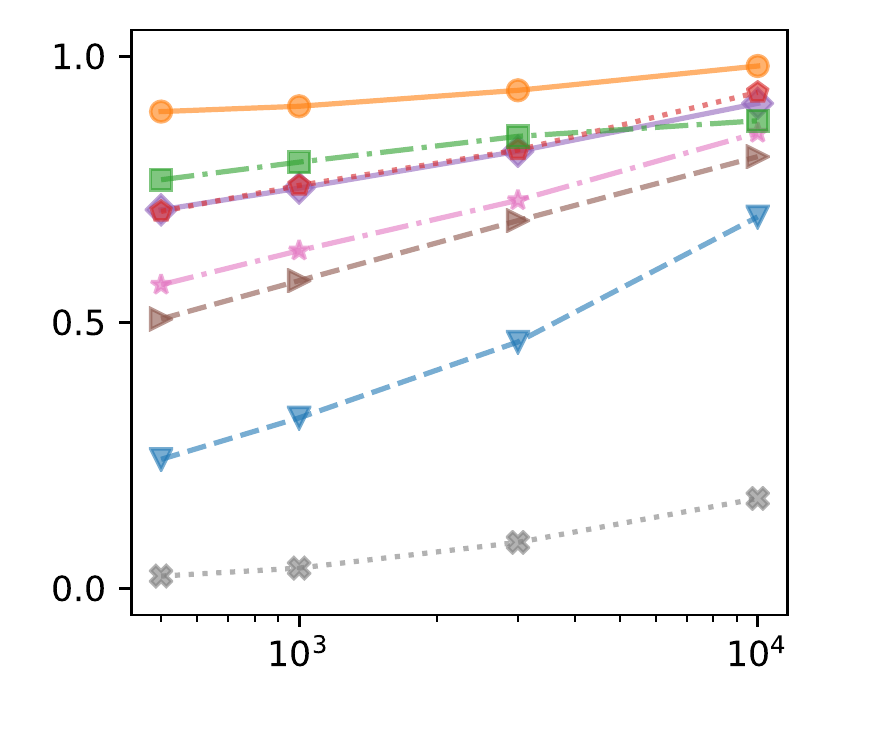} & \includegraphics[width=0.3\textwidth]{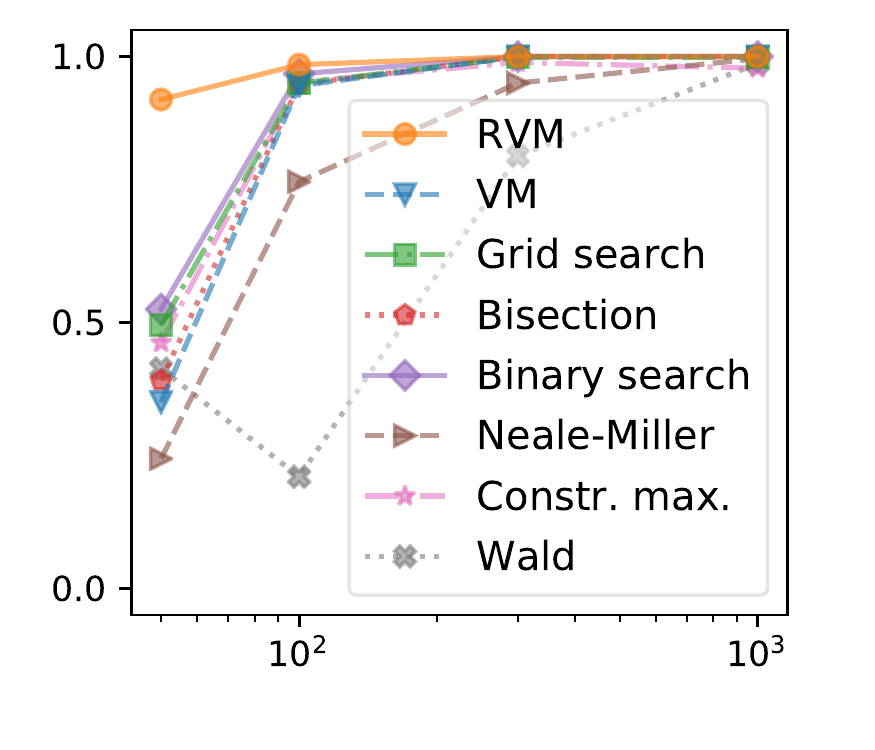}\tabularnewline
\begin{turn}{90}
\noindent \raggedleft{}\textsf{\footnotesize{}\hspace*{1cm}}%
\begin{minipage}[t]{3cm}%
\begin{center}
\textsf{\footnotesize{}Mean error}
\par\end{center}%
\end{minipage}
\end{turn} & \includegraphics[width=0.3\textwidth]{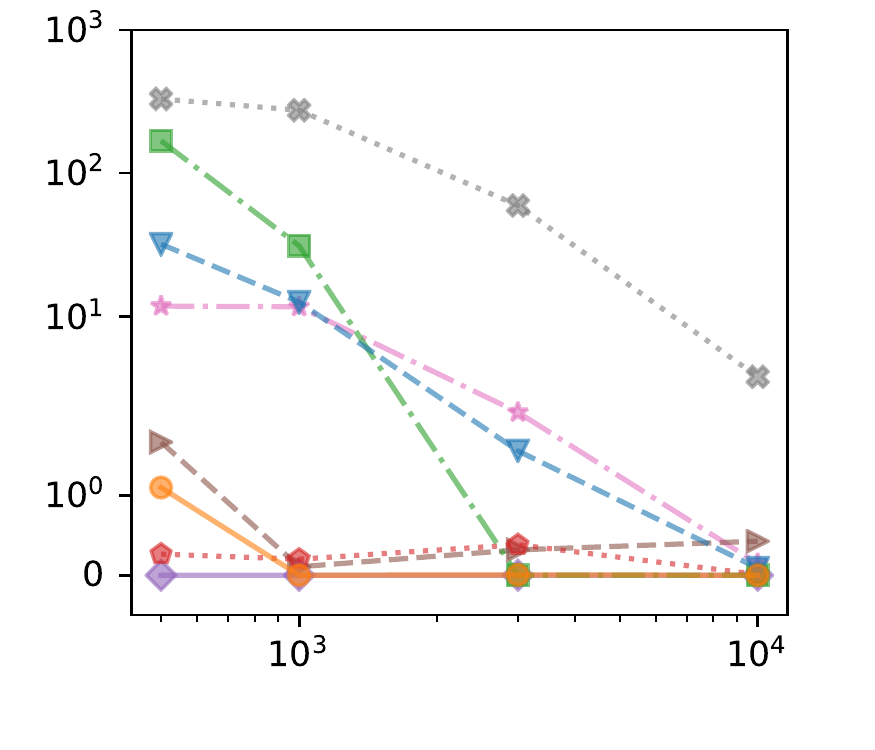} & \includegraphics[width=0.3\textwidth]{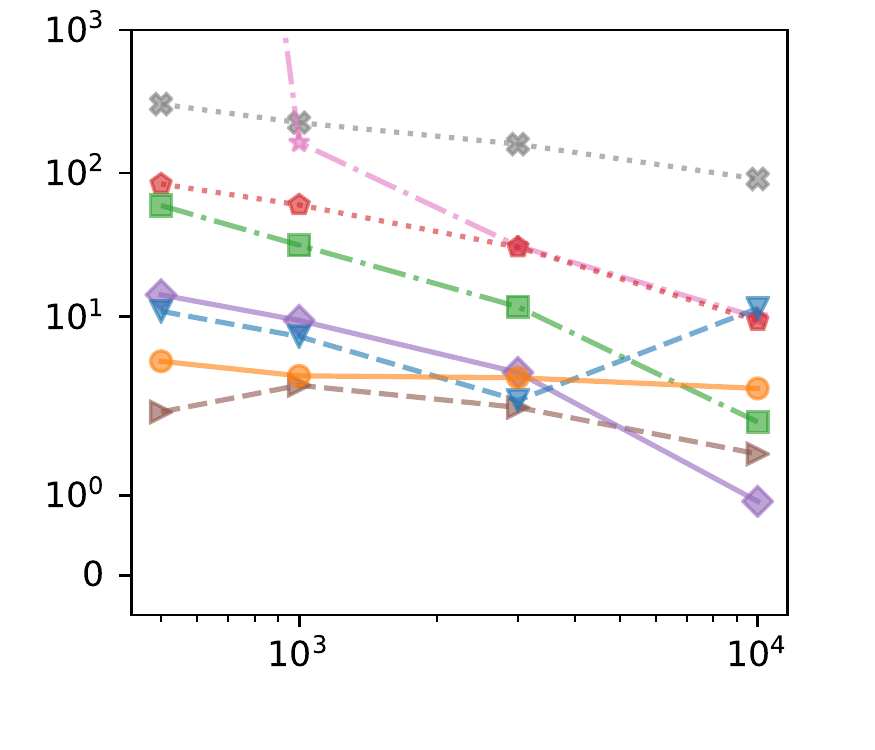} & \includegraphics[width=0.3\textwidth]{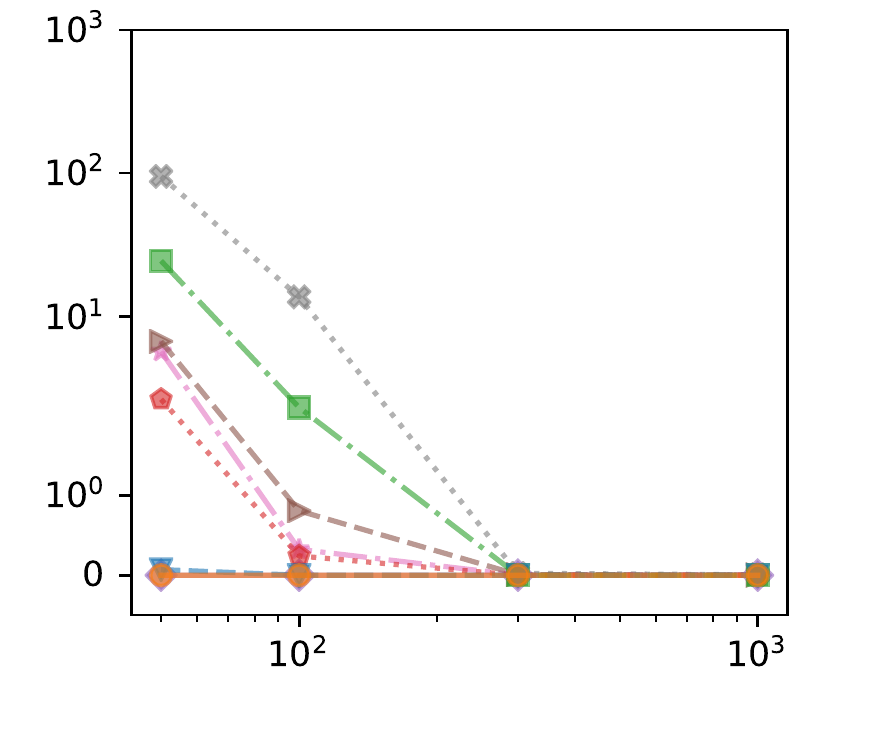}\tabularnewline
\begin{turn}{90}
\noindent \raggedleft{}\textsf{\footnotesize{}\hspace*{0.5cm}}%
\begin{minipage}[t]{4cm}%
\begin{center}
\textsf{\footnotesize{}Function evaluations}
\par\end{center}%
\end{minipage}
\end{turn} & \includegraphics[width=0.3\textwidth]{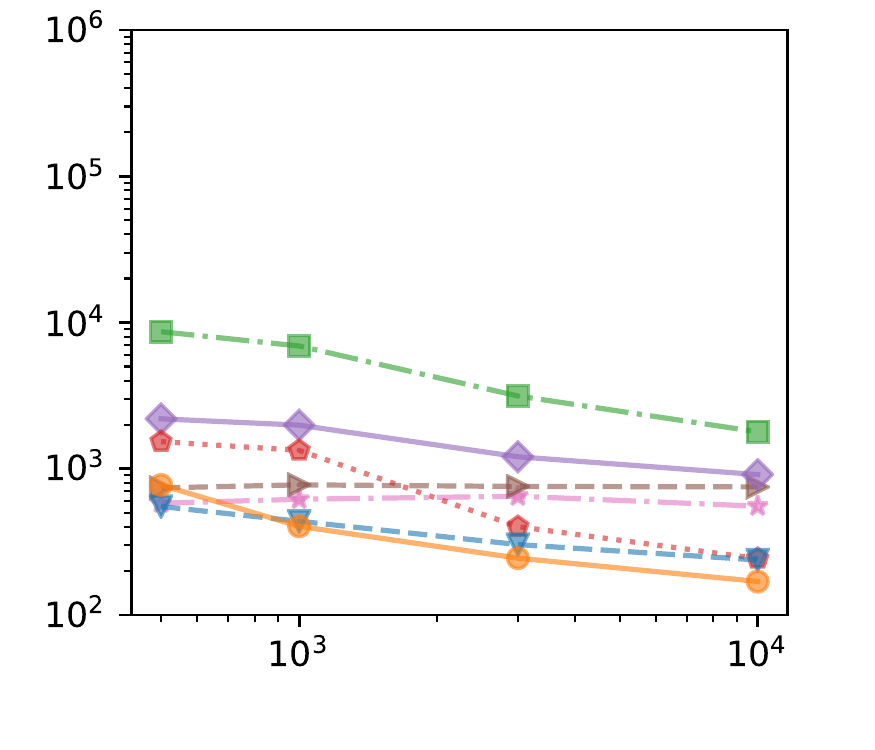} & \includegraphics[width=0.3\textwidth]{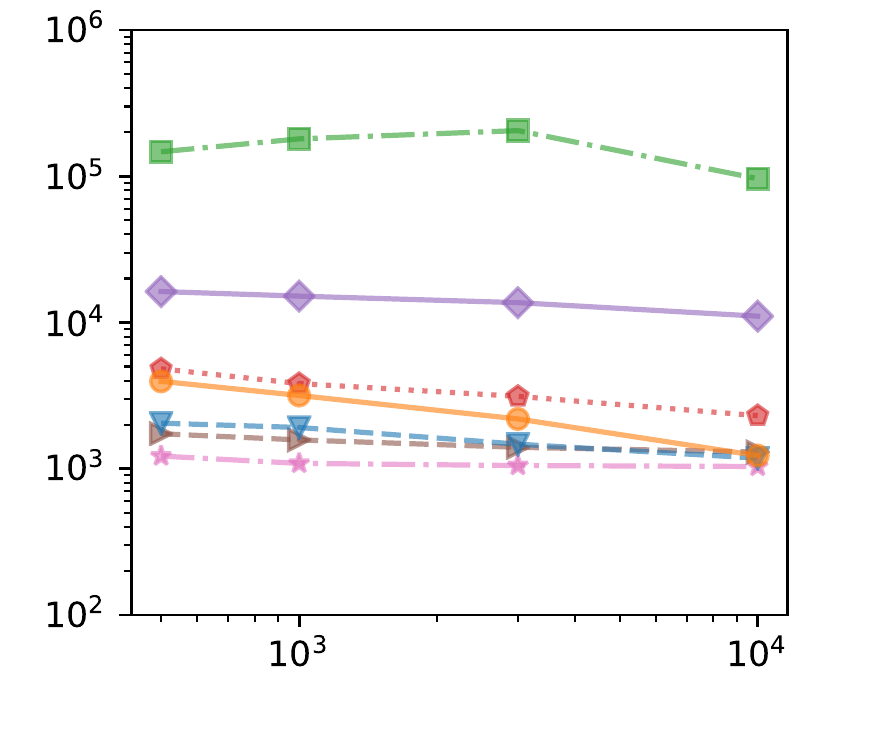} & \includegraphics[width=0.3\textwidth]{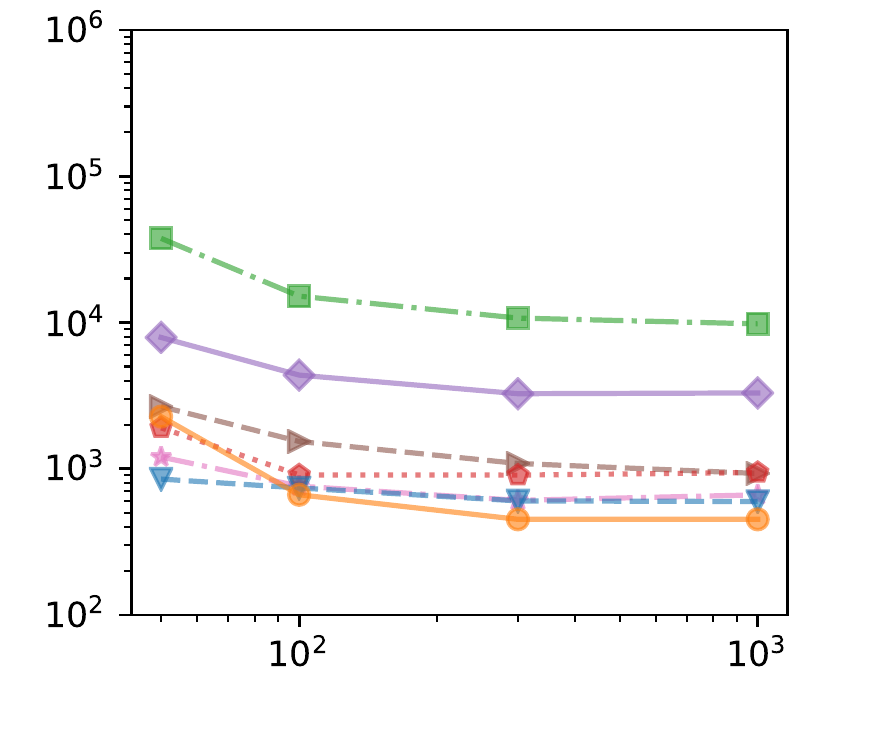}\tabularnewline
\begin{turn}{90}
\end{turn} & \centering{}\textsf{\footnotesize{}\hspace*{0.6cm}}%
\begin{minipage}[t]{4cm}%
\begin{center}
\textsf{\footnotesize{}$3$ parameters,}\\
\textsf{\footnotesize{}transformed covariates}
\par\end{center}%
\end{minipage} & \centering{}\textsf{\footnotesize{}\hspace*{0.7cm}}%
\begin{minipage}[t]{4cm}%
\begin{center}
\textsf{\footnotesize{}$11$ parameters, transformed covariates}
\par\end{center}%
\end{minipage} & \centering{}\textsf{\footnotesize{}\hspace*{0.6cm}$11$ parameters,
GLM}\tabularnewline
\end{tabular}

\caption[Benchmark results]{Benchmark results. The success rate, the mean error, and the number
of function evaluations are plotted for the $3$ parameter and the
$11$ parameter model with transformed covariates and for the $11$
parameter GLM. Throughout the simulations, our algorithm RVM had the
highest success rate. At the same time, RVM had a low mean error and
required only few likelihood function evaluations compared to the
considered alternative methods. The parameter values used to generate
the Figures are given in Supplementary Appendix B.\label{fig:Test-results-1}}
\end{figure*}

In fact, for all considered quality measures, RVM yielded good and
often the best results compared to the alternative methods (see Figure
\ref{fig:Test-results-1}). In all considered scenarios, RVM was the
algorithm with the highest success rate, which never fell below $90\%$
(second best: binary search, $52\%$). In scenarios with small data
sets, the success rate of RVM was up to $37$ percent points higher
than any other method. At the same time, RVM was among the fastest
algorithms. In scenarios with large data sets, RVM often converged
within three iterations. Furthermore, RVM was quick in the $3$ parameter
model, in which the Hessian matrix is easy to compute. In the scenario
with transformed covariates and $11$ parameters, RVM required about
three times as many likelihood evaluations as the fastest algorithm
but had a more than $56\%$ higher success rate. The error in the
results returned by RVM was consistently low compared to other methods.
The proportion of large errors was always below $1\%$, and the mean
error excluding these outliers never exceeded $0.05$.

The algorithms that require repeated evaluations of the profile likelihood
function performed second best in terms of the success rate. Except
for the GLM with $50$ data points, the binary search, the grid search,
and the bisection method consistently had success rates above $70\%$,
whereby the success rate increased with the size of the considered
data set. However, these algorithms also required more function evaluations
than other methods. In fact, the grid search was more than $5$ times
slower than any other algorithm. The binary search was slightly less
efficient than the bisection method, which exploits the approximately
quadratic shape of the profile likelihood function if many data are
available. In scenarios with large data sets, the bisection method
was among the most efficient algorithms. The errors of the three root
finding methods decreased the more data became available to fit the
models. However, while the binary search had a consistently low error,
both the grid search and the bisection method were more prone to large
errors than all other tested methods.

The algorithms developed from the constrained maximization perspective
(the method by \citeauthor{neale_use_1997} and direct constrained
maximization) had success rates ranging between $45\%$ and $85\%$
in problems with transformed covariates. In the GLM scenario, the
success rate was smaller in with $50$ data points and higher with
more data. The constrained maximization procedure was slightly more
successful than the method by \citet{neale_use_1997}. Both methods
required relatively few function evaluations, whereby direct constrained
maximization performed better. Both methods were less prone to large
errors than the grid search and the bisection method. However, the
outlier-reduced error was on average more than twice as large than
with any other method except RVM (\citeauthor{neale_use_1997}: $0.16$,
constrained maximum $0.09$, RVM: $0.07$).

The success of the algorithm VM depended highly on the properties
of the likelihood function. In scenarios with few data and transformed
covariates, VM had very low success rates (as low as $10\%$). When
more data were added, VM became as successful as the method by \citeauthor{neale_use_1997}
and direct constrained maximization. Thereby, VM was highly efficient
whenever results were obtained successfully. Similar to the success
rate, the mean error of VM decreased strongly as more data were considered.

Wald's method had very low success rates and large errors except for
the GLM with large data sets. In the models with transformed covariates,
Wald's method never had a success rate above $17\%$.

\section{Discussion\label{sec:Discussion-1}}

We presented an algorithm that determines the end points of profile
likelihood confidence intervals both of parameters and functions of
parameters with high robustness and efficiency. We tested the algorithm
in scenarios varying in parameter number, size of the data set, and
complexity of the likelihood function. In the tests, our algorithm
RVM was more robust than any other considered method. At the same
time, RVM was among the fastest algorithms in most scenarios. This
is remarkable, because there is typically a trade-off between robustness
and computational speed of optimization algorithms. RVM achieves this
result by exploiting the approximately quadratic form of the log-likelihood
surface in ``benign'' cases while maintaining high robustness with
the trust-region approach. Consequently, RVM naturally extends the
algorithm VM \citep{venzon_method_1988}, which appeared to be highly
efficient but lacking robustness in our tests.

Surprisingly, RVM turned out to be even more robust than methods based
on repeated evaluations of the profile likelihood. For the bisection
method and the binary search, this may be due to failures of internal
optimization routines, as initial guesses far from the solution can
hinder accurate convergence. The grid search method, in turn, was
often aborted due to the limited step size, which precluded the method
from identifying confidence bounds farther than $40$ units away from
the respective MLE. This, however, does not explain the comparatively
high error in the results of the grid search, as only successful runs
were considered. We therefore hypothesize that internal optimization
issues were responsible for some failures.

As expected, the algorithms that searched for the confidence interval
end points directly were more efficient but less robust than algorithms
that repeatedly evaluate the profile likelihood. Remarkably, a ``standard''
algorithm for constrained optimization performed slightly better than
an unconstrained optimizer operating on the modified target function
suggested by \citet{neale_use_1997}. This indicates that the approximation
introduced by \citet{neale_use_1997} might not be necessary and even
of disadvantage.

All methods implemented in this study (except RVM and VM) rely on
general optimizers. Consequently, the performance of these methods
depends on the chosen optimizers both in terms of computational speed
and robustness. Careful adjustment of optimization parameters might
make some of the implemented algorithms more efficient and thus more
competitive in benchmark tests. Though we attempted to reduce potential
bias by applying a variety of different methods, an exhaustive test
of optimization routines was beyond the scope of this study. Nonetheless,
the consistently good performance of RVM throughout our benchmark
tests suggests that RVM is a good choice in many applications.

Though RVM performed well in our tests, there are instances in which
the algorithm is not applicable or sufficiently efficient. This are
scenarios in which (1) the log-likelihood cannot be computed directly,
(2) the Hessian matrix of the log-likelihood function is hard to compute,
(3) the dimension of the parameter space is very large, or (4) there
are multiple points in the parameter space in which problem (\ref{eq:constrained-max})
is solved locally. Below, we briefly discuss each of these limitations.

(1) In hierarchical models, the likelihood function may not be known.
As RVM needs to evaluate the log-likelihood, its gradient, and its
Hessian matrix, the algorithm is not applicable in these instances.
Consequently, sampling based methods, such as parametric bootstrap
\citep{efron_nonparametric_1981}, Monte Carlo methods \citep{buckland_monte_1984},
or data cloning \citep{ponciano_hierarchical_2009} may then be the
only feasible method to determine confidence intervals.

(2) Especially in problems with a large parameter space, it is computationally
expensive to compute the Hessian matrix with finite difference methods,
as the number of function calls increases in quadratic order with
the length of the parameter vector. Though alternative differentiation
methods, such as analytical or automatic differentiation \citep{griewank_automatic_1989},
are often applicable, there may be some instances in which finite
difference methods are the only feasible alternative. In these scenarios,
optimization routines that do not require knowledge of the Hessian
matrix may be faster than RVM. Note, however, that the higher computational
speed may come with decreased robustness, and sampling based methods
might be the only remaining option if application of RVM is infeasible.

(3) If the parameter space has a very high dimension (exceeding $1000$),
internal routines, such as inversion of the Hessian matrix, may become
the dominant factor determining the speed of RVM. Though it may be
possible in future to make RVM more efficient, sampling based methods
or algorithms that do not use the Hessian matrix may be better suited
in these scenarios.

(4) RVM as well as all other methods implemented in this study are
local optimization algorithms. Therefore, the algorithms may converge
to wrong results if maximization problem (\ref{eq:constrained-max})
has multiple local solutions. This is in particular the case if the
confidence set  $\left\{ \th_{0}:\pl\ap{\th_{0}}\geq\ell^{*}\right\} $
is not connected and thus no interval. RVM reduces the issue of local
extreme points by choosing steps carefully and ensuring that the point
of convergence is indeed a maximum. This contrasts with VM, which
could converge to the wrong confidence interval end point (e.g. maximum
instead of minimum) if the initial guesses are not chosen with care.
Nonetheless, stochastic optimization routines, such as genetic algorithms
\citep{akrami_profile_2010}, and sampling methods may be better suited
if a local search is insufficient.

Despite these caveats, RVM is applicable to a broad class of systems.
Especially when inestimable parameters are present, commonly used
methods such as VM or grid search techniques can break down or be
highly inefficient. Furthermore, optimization failures are commonly
observed if not enough data are available to reach the asymptotic
properties of the MLE \citep{ren_algorithm_2019}. RVM is a particularly
valuable tool in these instances.

\section{Conclusion}

We developed and presented an algorithm to determine profile likelihood
confidence intervals. In contrast to many earlier methods, our algorithm
is robust in scenarios in which lack of data or a complicated likelihood
function make it difficult to find the bounds of profile likelihood
confidence intervals. In particular, our methods is applicable in
instances in which parameters are not estimable and in cases in which
the likelihood function has strong nonlinearities. At the same time,
our method efficiently exploits the asymptotic properties of the maximum
likelihood estimator if enough data are available.

We tested our method on benchmark problems with different difficulty.
Throughout our simulations, our method was the most robust while also
being among the fastest algorithms. We therefore believe that RVM
can be helpful to researchers and modelers across disciplines.

\bibliographystyle{apalike}
\bibliography{bibliography}

\end{document}


\sloppy
\global\long\def\argparentheses#1{\mathopen{\left(#1\right)}\mathclose{}}%
\global\long\def\ap#1{\argparentheses{#1}\mathclose{}}%

\global\long\def\var#1{\mathbb{V}\argparentheses{#1}}%
\global\long\def\pr#1{\mathbb{P}\argparentheses{#1}}%
\global\long\def\ev#1{\mathbb{E}\argparentheses{#1}}%

\global\long\def\smft#1#2{\stackrel[#1]{#2}{\sum}}%
\global\long\def\smo#1{\underset{#1}{\sum}}%
\global\long\def\prft#1#2{\stackrel[#1]{#2}{\prod}}%
\global\long\def\pro#1{\underset{#1}{\prod}}%
\global\long\def\uno#1{\underset{#1}{\bigcup}}%

\global\long\def\order#1{\mathcal{O}\argparentheses{#1}}%
\global\long\def\R{\mathbb{R}}%
\global\long\def\Q{\mathbb{Q}}%
\global\long\def\N{\mathbb{N}}%
\global\long\def\F{\mathcal{F}}%

\global\long\def\mathtext#1{\mathrm{#1}}%

\global\long\def\maxo#1{\underset{#1}{\max\,}}%
\global\long\def\argmaxo#1{\underset{#1}{\mathtext{argmax}\,}}%
\global\long\def\argsupo#1{\underset{#1}{\mathtext{argsup}\,}}%
\global\long\def\supo#1{\underset{#1}{\sup\,}}%
\global\long\def\info#1{\underset{#1}{\inf\,}}%
\global\long\def\mino#1{\underset{#1}{\min\,}}%
\global\long\def\argmino#1{\underset{#1}{\mathtext{argmin}\,}}%
\global\long\def\limo#1#2{\underset{#1\rightarrow#2}{\lim}}%
\global\long\def\supo#1{\underset{#1}{\sup}}%
\global\long\def\info#1{\underset{#1}{\inf}}%

\global\long\def\b#1{\boldsymbol{#1}}%
\global\long\def\ol#1{\overline{#1}}%
\global\long\def\ul#1{\underline{#1}}%

\newcommandx\der[3][usedefault, addprefix=\global, 1=, 2=, 3=]{\frac{d^{#2}#3}{d#1^{#2}}}%
\newcommandx\pder[3][usedefault, addprefix=\global, 1=, 2=]{\frac{\partial^{#2}#3}{\partial#1^{#2}}}%
\global\long\def\intft#1#2#3#4{\int\limits _{#1}^{#2}#3d#4}%
\global\long\def\into#1#2#3{\underset{#1}{\int}#2d#3}%

\global\long\def\th{\theta}%
\global\long\def\la#1{~#1~}%
\global\long\def\laq{\la =}%
\global\long\def\normal#1#2{\mathcal{N}\argparentheses{#1,\,#2}}%
\global\long\def\uniform#1#2{\mathcal{U}\argparentheses{#1,\,#2}}%
\global\long\def\I#1#2{\mbox{I}_{#1}\argparentheses{#2}}%
\global\long\def\chisq#1{\chi_{#1}^{2}}%
\global\long\def\dar{\,\Longrightarrow\,}%
\global\long\def\dal{\,\Longleftarrow\,}%
\global\long\def\dad{\,\Longleftrightarrow\,}%
\global\long\def\norm#1{\left\Vert #1\right\Vert }%
\global\long\def\code#1{\mathtt{#1}}%
\global\long\def\descr#1#2{\underset{#2}{\underbrace{#1}}}%
\global\long\def\NB{\mathcal{NB}}%
\global\long\def\BNB{\mathcal{BNB}}%
\global\long\def\e#1{\text{e}_{#1}}%
\global\long\def\P{\mathbb{P}}%
\global\long\def\pb#1{\Bigg(#1\Bigg)}%
\global\long\def\cpb#1{\Big[#1\Big]}%

\global\long\def\vv#1{\boldsymbol{#1}}%

\global\long\def\mt#1{\mathtext{#1}}%

\global\long\def\mat#1{\mt{#1}}%

\global\long\def\mm#1{\underline{\boldsymbol{#1}}}%

\global\long\def\G{g}%

\global\long\def\Gv{\vv{\G}}%

\global\long\def\H{\mat H}%

\global\long\def\Hv{\vv{\mat{\H}}}%

\global\long\def\Hm{\mm{\H}}%

\global\long\def\pl{\ell_{\mt{PL}}}%

\global\long\def\dx{\vv{\delta}}%
\global\long\def\dxt{\widetilde{\dx}}%
\global\long\def\dxn{\delta}%
\global\long\def\wt#1{\widetilde{#1}}%

\title{Supplementary Appendices for ``A Robust and Efficient Algorithm to
Find Profile Likelihood Confidence Intervals'' }

\author{Samuel M. Fischer         \and
		Mark A. Lewis
}
\institute{\email{samuel.fischer@ualberta.ca}}
\date{}

\maketitle

\appendix

\renewcommand{\theequation}{A\arabic{equation}}
\renewcommand{\thefigure}{A\arabic{figure}}
\renewcommand{\thetable}{A\arabic{table}}
\setcounter{figure}{0}
\setcounter{equation}{0}

\vspace*{-1.6cm}

\section{An alternative way to account for singular matrices\label{sec:Alternative-Singular}}

In each iteration, we seek to maximize the approximate likelihood
$\hat{\ell}$ with respect to the nuisance parameters. To that end,
we solve the equation 
\begin{equation}
0\laq\pder[\dxt]{}\hat{\ell}^{\dx}\laq\widetilde{\Hm}\wt{\dx^{*}}+\widetilde{\Hv}_{0}\dxn_{0}^{*}+\widetilde{\Gv}\label{eq:nuisance_grad}
\end{equation}
which has a unique solution if and only if $\widetilde{\Hm}$ is invertible.
Otherwise, equation (\ref{eq:nuisance_grad}) may have infinitely
or no solutions. In the main text, we suggested to solve (\ref{eq:nuisance_grad})
with the Moore-Penrose inverse if $\widetilde{\Hm}$ is singular.
However, this procedure appeared to be very sensitive to a threshold
parameter in tests, and we obtained better results with an alternative
method, which we describe below. We furthermore show test results
comparing the two methods.

\subsection{Description of the method}

If $\hat{\ell}$ has infinitely many maxima in the nuisance parameters,
we can choose some nuisance parameters freely and consider a reduced
system including the remaining independent parameters only. To that
end, we check $\widetilde{\Hm}$ for linear dependencies at the beginning
of each iteration. We are interested in a minimal set $S$ containing
indices of rows and columns whose removal from $\widetilde{\Hm}$
would make the matrix invertible. To compute $S$, we iteratively
determine the ranks of sub-matrices of $\widetilde{\Hm}$ using singular
value decompositions (SVMs). SVMs are a well-known tool to identify
the rank of a matrix.

The iterative algorithm proceeds as follows: first, we consider one
row of $\widetilde{\Hm}$ and determine its rank. Then, we continue
by adding a second row, determine the rank of the new matrix and repeat
the procedure until all rows, i.e. the full matrix $\widetilde{\Hm}$,
are considered. Whenever the matrix rank increases with addition of
a row, this row is linearly independent from the previous rows. Conversely,
the rows that do not increase the matrix rank are linearly dependent
on other rows of $\widetilde{\Hm}$. The indices of these rows form
the set $S$. 

In general, the set of linearly dependent rows is not unique. Therefore,
we consider the rows of $\widetilde{\Hm}$ in descending order of
the magnitudes of the corresponding gradient entries. This can help
the algorithm to converge faster. 

After $S$ is determined, we need to check whether there is a parameter
vector $\vv{\th}^{*}$ satisfying requirements 1 and 2 from section
2.1 (main text) for the approximation $\hat{\ell}$.
Let $\widetilde{\Hm}_{\mt{dd}}$ (``d'' for ``dependent'') be
the submatrix of $\Hm$ that remains if all rows and columns corresponding
to indices in $S$ are removed from $\widetilde{\Hm}$. Similarly,
let $\widetilde{\Hm}_{\mt{ff}}$ (``f'' for ``free'') be the submatrix
of $\widetilde{\Hm}$ containing only the rows and columns corresponding
to indices in $S$, and let $\widetilde{\Hm}_{\mt{df}}=\widetilde{\Hm}_{\mt{fd}}^{\top}$
be the matrix containing the rows whose indices are \emph{not }in
$S$ and the columns whose indices \emph{are} in $S$. Let us define
$\widetilde{\Gv}_{\mt d}$, $\widetilde{\Gv}_{\mt f}$, $\widetilde{\dx}_{\mt d}$,
and $\widetilde{\dx}_{\mt f}$ accordingly. If $\widetilde{\Hm}_{\mt{dd}}$
is not negative definite, $\hat{\ell}$ is unbounded, and requirement
2 (section 2.1 main text) cannot be satisfied. Otherwise, we may attempt
to solve 
\begin{eqnarray}
0 & = & \pder[\dxt]{}\hat{\ell}^{\dx}\\
\dad\nonumber \\
0 & = & \widetilde{\Hm}_{\mt{dd}}\wt{\dx_{\mt d}^{*}}+\widetilde{\Hm}_{\mt{df}}\wt{\dx_{\mt f}^{*}}+\widetilde{\Hv}_{0\mt d}\dxn_{0}^{*}+\widetilde{\Gv}_{\mt d}\label{eq:nuisance-system-lindep-indep-1}\\
0 & = & \widetilde{\Hm}_{\mt{df}}^{\top}\wt{\dx_{\mt d}^{*}}+\widetilde{\Hm}_{\mt{ff}}\wt{\dx_{\mt f}^{*}}+\widetilde{\Hv}_{0\mt f}\dxn_{0}^{*}+\widetilde{\Gv}_{\mt f}.\label{eq:nuisance-system-lindep-dep-1}
\end{eqnarray}
If equation system (\ref{eq:nuisance-system-lindep-indep-1})-(\ref{eq:nuisance-system-lindep-dep-1})
has a solution, we can choose $\wt{\dx_{\mt f}^{*}}$ freely. Setting
$\wt{\dx_{\mt f}^{*}}\leftarrow\vv 0$ makes equation (\ref{eq:nuisance-system-lindep-indep-1})
equivalent to 
\begin{eqnarray}
\wt{\dx_{d}^{*}} & = & -\widetilde{\Hm}_{\mt{dd}}^{-1}\left(\widetilde{\Hm}_{0\mt d}\dxn_{0}+\widetilde{\Gv}_{\mt d}\right).
\end{eqnarray}
That is, we may set $\widetilde{\Hm}\leftarrow\widetilde{\Hm}_{\mt{dd}}$,
$\widetilde{\G}\leftarrow\widetilde{\G}_{\mt d}$, $\widetilde{\dx^{*}}\leftarrow\wt{\dx_{\mt d}^{*}}$
for the remainder of the current iteration and proceed as usual, whereby
the free nuisance parameters are left unchanged: $\wt{\dx_{\mt f}^{*}}=\vv 0$.
With the resulting $\dxn_{0}^{*}$, we check whether (\ref{eq:nuisance-system-lindep-dep-1})
holds approximately. If not, the log-likelihood is unbounded above.
We consider this case in section 2.4 in the main text.

\subsection{Tests}

\begin{figure*}[t]
	\begin{tabular}[t]{>{\raggedleft}p{0.5cm}>{\centering}p{0.27\textwidth}>{\centering}p{0.27\textwidth}>{\centering}p{0.27\textwidth}}
		\begin{turn}{90}
		\end{turn} & \centering{}\textsf{\footnotesize{}\hspace*{0.8cm}}%
		\begin{minipage}[t]{3cm}%
			\begin{center}
				\textsf{\footnotesize{}A}
				\par\end{center}%
		\end{minipage} & \centering{}\textsf{\footnotesize{}\hspace*{0.8cm}}%
		\begin{minipage}[t]{3cm}%
			\begin{center}
				\textsf{\footnotesize{}B}
				\par\end{center}%
		\end{minipage} & \centering{}\textsf{\footnotesize{}\hspace*{0.8cm}}%
		\begin{minipage}[t]{3cm}%
			\begin{center}
				\textsf{\footnotesize{}C}
				\par\end{center}%
		\end{minipage}\tabularnewline
		\begin{turn}{90}
			\textsf{\footnotesize{}\hspace*{1cm}}%
			\begin{minipage}[t]{3cm}%
				\begin{center}
					\textsf{\footnotesize{}Success rate}
					\par\end{center}%
			\end{minipage}
		\end{turn} & \includegraphics[width=0.3\textwidth]{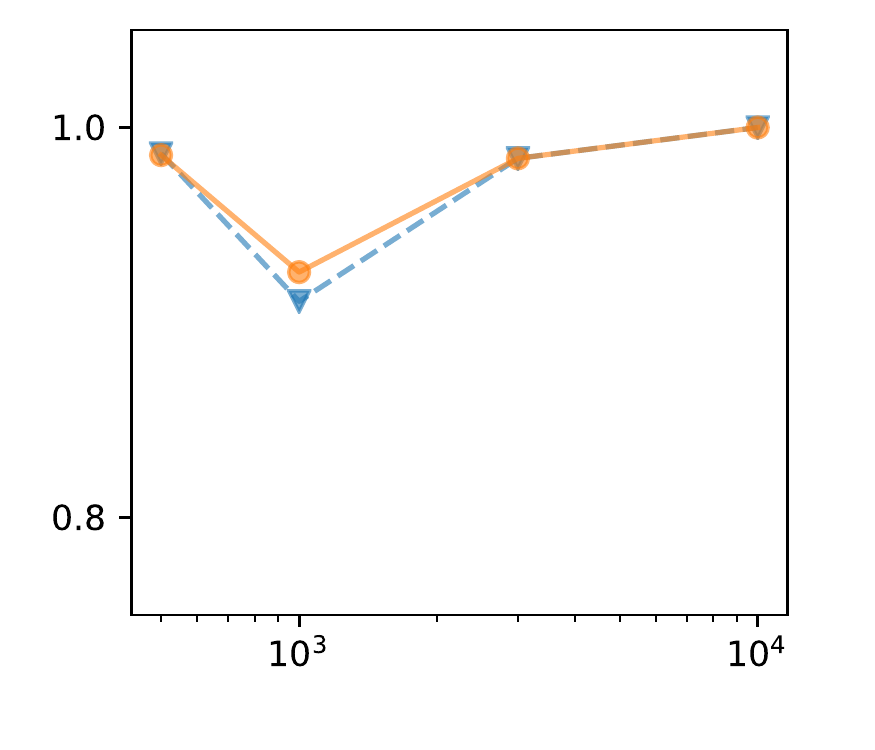} & \includegraphics[width=0.3\textwidth]{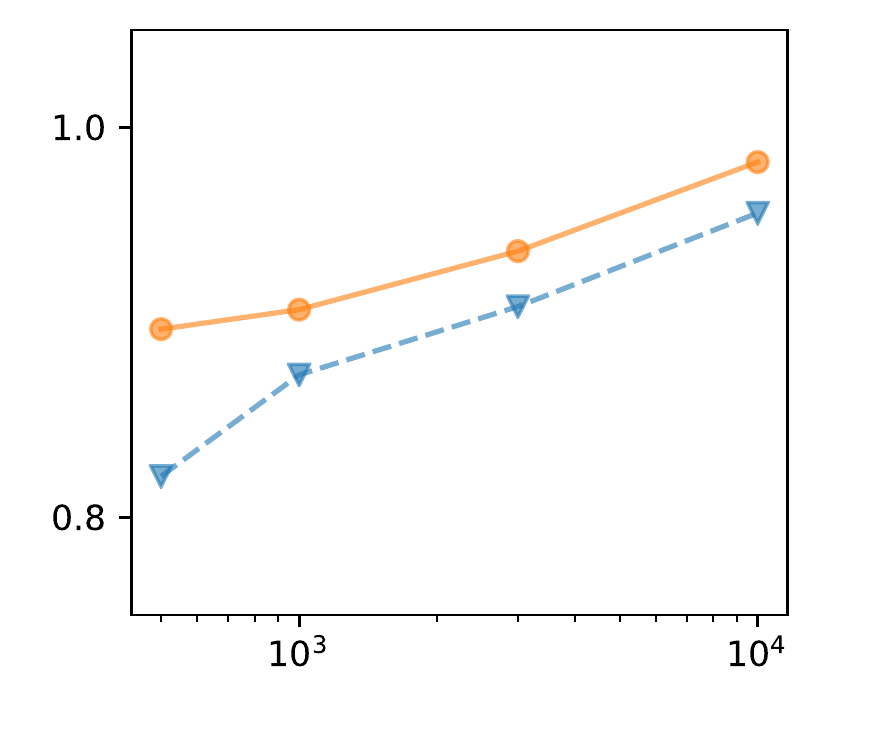} & \includegraphics[width=0.3\textwidth]{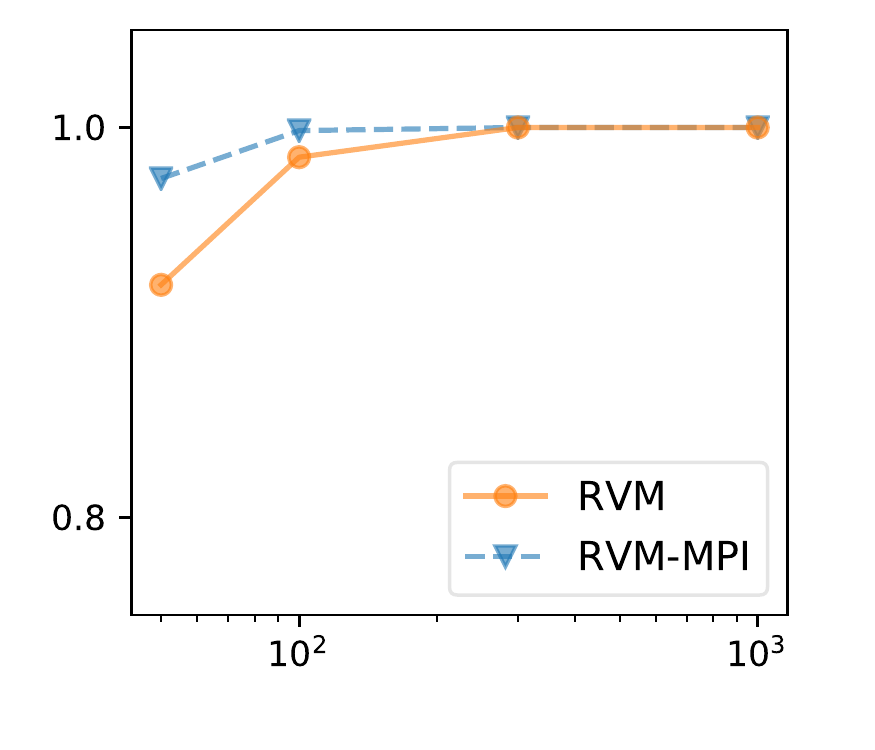}\tabularnewline
		\begin{turn}{90}
		\end{turn} & \centering{}\textsf{\footnotesize{}\hspace*{0.6cm}}%
		\begin{minipage}[t]{4cm}%
			\begin{center}
				\textsf{\footnotesize{}$3$ parameters,}\\
				\textsf{\footnotesize{}transformed covariates}
				\par\end{center}%
		\end{minipage} & \centering{}\textsf{\footnotesize{}\hspace*{0.7cm}}%
		\begin{minipage}[t]{4cm}%
			\begin{center}
				\textsf{\footnotesize{}$11$ parameters, transformed covariates}
				\par\end{center}%
		\end{minipage} & \centering{}\textsf{\footnotesize{}\hspace*{0.6cm}$11$ parameters,
			GLM}\tabularnewline
	\end{tabular}
	
	\caption[Comparison of different methods to handle linearly dependent parameters]{Comparison of different methods to handle linearly dependent parameters.
		The success rate of RVM is plotted for the $3$ parameter and the
		$11$ parameter model with transformed covariates and for the $11$
		parameter GLM. Though the algorithm using the Moore-Penrose inverse
		(RVM-MPI) performed slightly better for the GLM with little data (panel
		C), the method had lower success rates when the $11$ parameter model
		with transformed variables was considered (panel B). The parameter
		values used to generate the Figures are given in Appendix \ref{sec:Parameters-for-benchmark-tests}.\label{fig:Test-results-2}}
\end{figure*}

We implemented RVM with both suggested methods for treating linearly
dependent parameters. To that end, we applied the same testing procedure
described in section 3 of the main text. The two methods yielded similar
results in terms of computational speed (number of required likelihood
evaluations) and error in case of reported success (see section 3.1
in the main text). However, holding some parameters constant as suggested
in this Appendix turned out to be more robust in general and lead
to slightly higher success rates (see Figure \ref{fig:Test-results-2}).
Therefore, we suggest using this method in practice.

\onecolumn

\section{Parameters for benchmark tests\label{sec:Parameters-for-benchmark-tests}}

Here we provide the parameter values used to generate the data for
our benchmark tests. We tested models with transformed covariates
with $3$ and $11$ parameters and a GLM with $11$ parameters. For
the models with transformed covariates, we considered scenarios with
$N=500$, $N=1000$, $N=3000$, and $N=10000$ data points. The parameters
for the two model families are given in Tables \ref{tab:3-parameters}
and \ref{tab:11-parameters}. For the GLM, we considered data sets
with sizes $N=50$, $N=100$, $N=300$, and $N=1000$. We provide
the parameter values in Table \ref{tab:11cx-parameters}.
\begin{table*}[h]
\begin{centering}
\footnotesize\renewcommand{\arraystretch}{1.5} \setlength\arrayrulewidth{0.5pt}
\par\end{centering}
\begin{raggedright}
\begin{tabular}{l>{\raggedright}p{1cm}>{\centering}p{1.2cm}>{\centering}p{1.2cm}}
\toprule 
\textbf{Parameter} & \centering{}$\alpha_{1}$ & \centering{}$\beta_{0}$ & \centering{}$\beta_{1}$\tabularnewline
\midrule 
\textbf{Value} & \centering{}$0.5$ & \centering{}$-10$ & \centering{}$5$\tabularnewline
\bottomrule
\end{tabular}
\par\end{raggedright}
\raggedright{}\caption[Parameters for the model with $3$ parameters and transformed covariates]{Parameters for the model with $3$ parameters and transformed covariates.
\label{tab:3-parameters}}
\end{table*}
\begin{table*}[h]
\begin{centering}
\footnotesize\renewcommand{\arraystretch}{1.5} \setlength\arrayrulewidth{0.5pt}
\par\end{centering}
\begin{raggedright}
\begin{tabular}{l>{\raggedright}p{0.8cm}>{\centering}p{0.8cm}>{\centering}p{0.8cm}>{\centering}p{0.8cm}>{\centering}p{0.8cm}>{\centering}p{0.8cm}>{\centering}p{0.8cm}>{\centering}p{0.8cm}>{\centering}p{0.8cm}>{\centering}p{0.8cm}>{\centering}p{0.8cm}}
\toprule 
\textbf{Parameter} & \centering{}$\alpha_{1}$ & \centering{}$\alpha_{2}$ & \centering{}$\alpha_{3}$ & \centering{}$\alpha_{4}$ & \centering{}$\alpha_{5}$ & \centering{}$\beta_{0}$ & \centering{}$\beta_{1}$ & \centering{}$\beta_{2}$ & \centering{}$\beta_{3}$ & \centering{}$\beta_{4}$ & \centering{}$\beta_{5}$\tabularnewline
\midrule 
\textbf{Value} & \centering{}$0.2$ & \centering{}$1$ & \centering{}$0.1$ & \centering{}$0.2$ & \centering{}$0.5$ & \centering{}$-1$ & \centering{}$5$ & \centering{}$2$ & \centering{}$-1$ & \centering{}$-3$ & \centering{}$-2$\tabularnewline
\bottomrule
\end{tabular}
\par\end{raggedright}
\caption[Parameters for the model with $11$ parameters and transformed covariates]{Parameters for the model with $11$ parameters and transformed covariates.
\label{tab:11-parameters}}
\end{table*}
\begin{table*}[h]
\begin{centering}
\footnotesize\renewcommand{\arraystretch}{1.5} \setlength\arrayrulewidth{0.5pt}
\par\end{centering}
\begin{raggedright}
\begin{tabular}{l>{\raggedright}p{0.8cm}>{\centering}p{0.8cm}>{\centering}p{0.8cm}>{\centering}p{0.8cm}>{\centering}p{0.8cm}>{\centering}p{0.8cm}>{\centering}p{0.8cm}>{\centering}p{0.8cm}>{\centering}p{0.8cm}>{\centering}p{0.8cm}>{\centering}p{0.8cm}}
\toprule 
\textbf{Parameter} & \centering{}$\beta_{0}$ & \centering{}$\beta_{1}$ & \centering{}$\beta_{2}$ & \centering{}$\beta_{3}$ & \centering{}$\beta_{4}$ & \centering{}$\beta_{5}$ & \centering{}$\beta_{6}$ & \centering{}$\beta_{7}$ & \centering{}$\beta_{8}$ & \centering{}$\beta_{9}$ & \centering{}$\beta_{10}$\tabularnewline
\midrule 
\textbf{Value} & \centering{}$0.8$ & \centering{}$0.2$ & \centering{}$-0.6$ & \centering{}$-1$ & \centering{}$-1$ & \centering{}$0.2$ & \centering{}$0.5$ & \centering{}$0.1$ & \centering{}$-0.2$ & \centering{}$0.2$ & \centering{}$2$\tabularnewline
\bottomrule
\end{tabular}
\par\end{raggedright}
\caption[Parameters for the $11$ parameter GLM]{Parameters for the $11$ parameter GLM. The covariate powers $\alpha_{i}$
are all fixed at $1$. \label{tab:11cx-parameters}}
\end{table*}